\documentclass[journal]{IEEEtran}
\input epsf
\usepackage{graphicx}
\usepackage{tabularx}
\usepackage{todonotes}

\ifx\pdfoutput\undefined
\usepackage[hypertex]{hyperref}
\else
\usepackage[pdftex,hypertexnames=false]{hyperref}
\fi
\hypersetup{
    colorlinks=true,
    linkcolor=blue,
    filecolor=magenta,      
    urlcolor=cyan,
    pdftitle={Overleaf Example},
    pdfpagemode=FullScreen,
}

\usepackage{caption}
\usepackage{subcaption}

\usepackage{algorithm}
\usepackage{algpseudocode}
\usepackage{amsmath}
\usepackage{caption} 
\algnewcommand\algorithmicforeach{\textbf{for each}}
\algdef{S}[FOR]{ForEach}[1]{\algorithmicforeach\ #1\ \algorithmicdo}
\usepackage{float}
\usepackage{booktabs}

\usepackage{cleveref}
\usepackage{balance}

\usepackage{xcolor}

\newcounter{reviewer}
\newcounter{comment}[reviewer]

\begin{document}
\title{Differential Area Analysis for Ransomware: Attacks, Countermeasures, and Limitations}
\author{Marco~Venturini,  
Francesco~Freda, 
Emanuele~Miotto, 
Mauro~Conti, 
and~Alberto~Giaretta
\thanks{M. Venturini, F. Freda, E. Miotto, and M. Conti,
are with the Department of Mathematics, University of Padova, Padova 35131, Italy}
\thanks{A. Giaretta is with the Department of Computer Science, \"Orebro University, \"Orebro 70281, Sweden}
\thanks{Manuscript received March 12 2023; revised September 20, 2024.}}


\maketitle

\begin{abstract}
Crypto-ransomware attacks have been a growing threat over the last few years. The goal of every ransomware strain is encrypting user data, such that attackers can later demand users a ransom for unlocking their data. To maximise their earning chances, attackers equip their ransomware with strong encryption which produce files with high entropy values. 
Davies et al. proposed Differential Area Analysis (DAA), a technique that analyses files headers to differentiate compressed, regularly encrypted, and ransomware-encrypted files.
In this paper, first we propose three different attacks to perform malicious header manipulation and bypass DAA detection. Then, we propose three countermeasures, namely 2-Fragments (2F), 3-Fragments (3F), and 4-Fragments (4F), which can be applied equally against each of the three attacks we propose. We conduct a number of experiments to analyse the ability of our countermeasures to detect ransomware-encrypted files, whether implementing our proposed attacks or not. Last, we test the robustness of our own countermeasures by analysing the performance, in terms of files per second analysed and resilience to extensive injection of low-entropy data. Our results show that our detection countermeasures are viable and deployable alternatives to DAA.
\end{abstract}
\IEEEoverridecommandlockouts

%
\IEEEpeerreviewmaketitle

\section{Introduction}
\label{sec:introduction}
Ransomware is a type of malware that encrypts victims' data with a strong encryption algorithm. Unless the victims pay the requested ransom, the attacker threatens to release or delete the locked data. The popularity of this attack reached the point that ransomware attacks are among the fastest growing threats in recent cybersecurity history. In 2021, SonicWall registered 623.3 million ransomware attacks worldwide, a 105\% increase over the attacks occurred in 2020, and more than triple the attacks in 2019~\cite{sonicwall2022}. Statistics gathered by Statista show that in 2020 alone, 127 new families of ransomware were discovered~\cite{statistafamilies}, and worldwide ransomware attacks grew to 304 millions, from 190 millions estimated in 2019~\cite{statistaattacks}.

Ransomware can have serious consequences for the victims it infects, with implications that transcend the digital world. Hospitals around the world have been regularly infected with ransomware and their regular operations have been severely impacted. In a French hospital, the attack was so disruptive that staff had to resort to pen and paper for weeks, and radiotherapy care was among the most affected units~\cite{portswigger}.

To detect ransomware infections and counteract their threat, security researchers proposed two different approaches. First, static analysis can be performed on software suspected of being ransomware before running it, to identify known sequences of bytes previously detected in other ransomware families. Although static analysis is fast and accurate on some families, evasive strains can avoid these countermeasures. One evasion technique relies on obfuscation techniques to change the structure of the ransomware itself~\cite{alrimy2018}. Another way to bypass static analysis is to implement polymorphism, as done by Reveton, WinLock, and Urausy ransomware families~\cite{kharraz2015}.

The second approach to ransomware detection is dynamic analysis, which consists of monitoring and analysing suspicious running processes to identify behavioural patterns that match known ransomware attacks. Capable of detecting evasive strains, dynamic analysis approaches have drawbacks as well. For example, they require to be sandboxed in a safe and monitored environment where the system can run them and analyse their behaviour. However, some ransomware families incorporate sandbox evasion techniques that allow them to understand whether they are being sandboxed (hence, analysed) or not.

As noted by Davies et al.~\cite{davies}, the state of the art for static and dynamic approaches shows low detection success rates; therefore, different approaches are necessary. A number of proposals stem from the observation that crypto-ransomware strains have a common denominator: at a certain point during their life-cycle, they will attempt to encrypt the victim's files. When a file undergoes encryption, its entropy naturally varies, and this variation can be considerable. As we cover in \Cref{sec:relatedworks}, various works
have proposed to monitor different changes in entropy to detect if files are currently being mass-encrypted, a telling sign of a potential ransomware infection. 
However, calculating the entropy over the entirety of a large set of files produces a relevant overhead to the system. In their work, Davies et al.~\cite{davies} monitor and analyse only the files headers, showing that it is possible to achieve a high success rate of detection while keeping the overhead acceptable.

Davies et al. proposal is effective and has solid roots on how ransomware currently operates and behaves. However, it is likely that ransomware will implement countermeasures to manipulate entropy and circumvent their detection strategy. In this paper, we show how ransomware could elude Davies et al. detection approach, and we propose some countermeasures to mitigate such adapting behaviour. In particular, the contributions of this paper are three-fold:
\begin{itemize}
\item We propose three attacks that ransomware could use to alter the files and reduce the detection success rate of Davies et al.'s DAA algorithm;
\item We suggest three different countermeasures for counteracting the attacks we propose;
\item We conduct thorough experiments to analyse the effectiveness of the proposed countermeasures against our attacks and against current ransomware operations. 
\end{itemize}

\subsection{Outline}
The paper is structured as follows: 
\Cref{sec:background} provides the background information necessary for the scope of this paper, while in \Cref{sec:relatedworks} we present an overview of the relevant related work. \Cref{sec:attacks} describes the attacks that we propose and shows the effectiveness against DAA detection. \Cref{sec:countermeasures} discusses three countermeasures against our attacks (2F, 3F, and 4F), complete with pseudocode algorithms, and \Cref{sec:count_results} shows the effectiveness of such countermeasures against our attacks, their accuracy when applied to the original DAA dataset, their time performance in terms of files per second analysed, and their robustness against further entropy manipulation. Last, in \Cref{sec:limitations} we discuss the limitations of our work and in \Cref{sec:conclusion} we draw our conclusions.

\section{Background}\label{sec:background}
In this section, we provide the background information that is necessary for the scope of our work. In particular, we provide a definition for the concept of entropy, and we recap the basic metrics that we use throughout the paper for evaluating our attacks and countermeasures.

\subsection{Entropy}
Entropy is defined as a measure of randomness or disorder. In information theory, the entropy of a variable is the average uncertainty of the possible outcomes of the variable itself. First defined by Claude Shannon~\cite{shannon1948}, given a discrete random variable $X$ that can assume values between $1$ to $n$ with their relatives probabilities of appearance $p(x_i)$, the Shannon entropy can be expressed using the following formula:
\begin{equation}
    H(X)=-\sum_{i=1}^n P(x_i)log_2 P(x_i).
\end{equation}

In terms of files, entropy can be seen as a measure of the predictability of the next files bytes, based on the previous ones. Therefore, a structured and regular file presents high predictability, hence low entropy, whereas a file composed of random values will exhibit high entropy.

\subsection{Evaluation metrics}
Throughout the paper, we use four standard metrics to evaluate the performance of our attacks and countermeasures. Namely, accuracy, precision, recall, and F1 score. Since these metric are well-established, we provide an intuitive description of their meaning in the scope of this research, without any standard formalisation.

\paragraph*{Accuracy} the total number of files that have been classified correctly.
    
\paragraph*{Precision} the ratio of the files correctly identified as encrypted by ransomware, among the total number of files classified as encrypted by ransomware.
    
\paragraph*{Recall} the ratio of the files correctly identified as encrypted by ransomware, among the total number of files truly encrypted by ransomware. 
    
\paragraph*{F1 Score} the harmonic mean of the Precision and Recall.

\subsection{DAA}\label{sec:daa}
\begin{figure*}
  \centering \includegraphics[width=\linewidth]{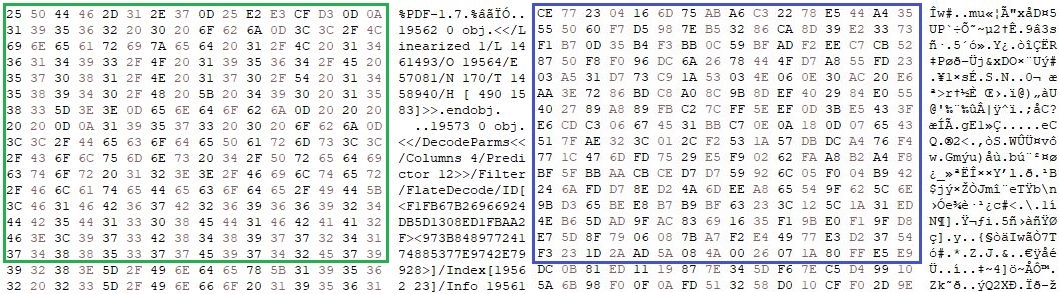}
  \caption{On the left, in the green box, the first 256 bytes of a legitimate PDF file. On the right, in the blue box, the first 256 bytes of a pseudo-random file.}
  \label{img:files_bytes}
\end{figure*}

Davies et al.'s DAA approach~\cite{davies}, used as a basis for this paper, works as follows. It takes the first 256 bytes of a file and assigns to each byte a value between 0 and 255 ($n=256$). The fundamental idea is to verify how the entropy varies with increased portions of the header. Therefore, the algorithm first calculates the entropy of the first 8 bytes of the fragment, then the entropy of the first 16 bytes, then the first 24 bytes, and it goes by 8 bytes increments until the full 256 bytes. To obtain a reference graph, the same algorithm is applied to a fragment containing 256 bytes of random data.
\Cref{img:files_bytes} shows the first 256 bytes of a PDF file circled in green, and the first 256 bytes of a random file circled in blue.

\begin{figure}
\centering
\includegraphics[trim={0 0 0 1.5em},clip,width=0.95\linewidth]{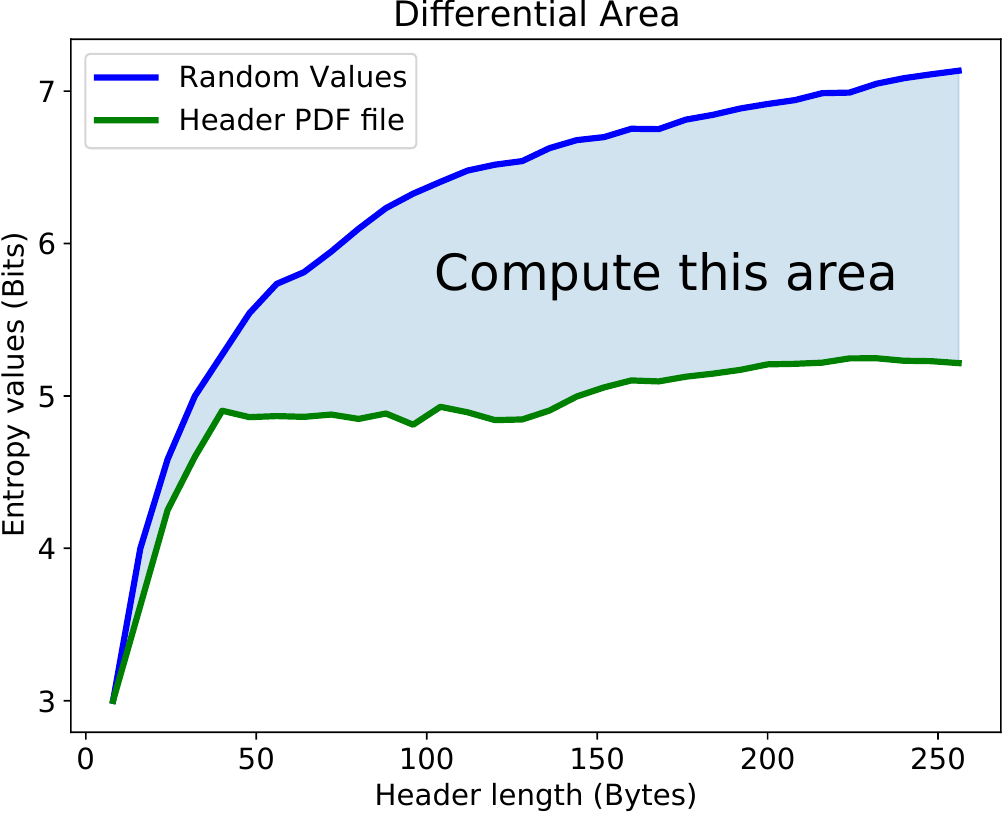}
\caption{The green curve shows the entropy analysis of a legitimate PDF file, performed in incremental steps, as done by Davies et al.~\cite{davies}. The blue curve shows the result of the same process performed on a pseudo-random file, similar to the ones produced by ransomware. The grey area between the two lines is the differential area, as proposed by Davies et al. with their DAA approach.}
\label{img:daa-plot}
\end{figure}

The two curves are then combined in a graph similar \Cref{img:daa-plot}, and the area between the two curves is calculated. For example, \Cref{img:daa-plot} shows in blue the curve obtained with 256 random bytes and in green the curve that results from the analysis of a PDF file header, and the differential area between the two lines (calculated using the Composite Trapezoidal Rule, as shown in \Cref{sec-mit-ov}) indicates to which degree the PDF file is random. If the analysed file has high entropy, the area is small, whereas if the file has low entropy, the area is large. The former case could indicate that the file has been encrypted by ransomware, and the latter case would indicate that the file is legitimate. In particular, the method calculates the difference between the file entropy and the random sequence entropy for each step of the algorithm (i.e., for [8, 16, 24, ..., 256] bytes), comparing the results to a predefined threshold. If the differential area is lower than the threshold at any step, the file contains too many random bytes for being legitimate, and the algorithm issues a warning.

\section{Related works}\label{sec:relatedworks}
The use of entropy for detecting malware infections has been investigated by several researches. In this section, we aim to provide an overview regarding the different uses of entropy in the ransomware detection state-of-the-art. 

First, it is possible to approach the problem of ransomware detection and mitigation by analysing the infecting binaries. Zhu et al.~\cite{zhu2022} noted that, while deep learning approaches would be helpful for analysing ransomware binaries and classifying them in classes, they are unsuitable when the training samples are limited. Therefore, they propose a few-shot meta-learning Siamese neural network that compares the entropy features of different ransomware, showing that they exhibit different entropy graphs across classes, but similar graphs within classes. Their experiments, conducted on 11 ransomware classes, achieved a weighted F1-score above 86\%, proving the feasibility of the proposed approach.

Hirano et al.~\cite{hirano2019}, proposed to detect zero-day ransomware attacks by applying machine learning to suspicious I/O requests. In their paper, they tested Random Forest, SVM, and K-nearest neighbours, using as features Shannon entropy and other information extracted from the collected traffic data. However, unlikely previous works in the literature, the authors adopt a window-based feature computation, preserving time-series characteristics in hardware-level storage access patterns. This allows to achieve a framework that is agnostic with respect to the operating system or kernel drivers.

Cuzzocrea et al.~\cite{cuzzocrea2018} used the entropy measure to produce a fuzzy logic algorithm for detecting suspicious Android applications. The authors tested eight fuzzy logic classifiers, concluding that the FuzzyRoughNN algorithm allows for achieving a weighted precision of 84\% and a weighted recall of 83\%. The results were obtained using a mixed dataset of 10000 applications, both legitimate and ransomware.

Some state-of-the-art use entropy to analyse processes behaviour over an extended span of time. For example, Continella et al.~\cite{continella} proposed ShieldFS, a ransomware detection system that analyses the historical entropy of various processes features, such as write operations, frequency of read, write, and folder-listing operations, dispersion of per-file writes, fraction of files renamed, and file-type usage statistics. This allows ShieldFS to distinguish ransomware from benign processes at runtime, enhancing the resilience of operating systems to ransomware infections and their short-term file modifications.

Other approaches to ransomware detection aim to detect suspicious modifications to users' files. In fact, Genç et al.~\cite{genc2018} listed the inflation of entropy after file-write operations as one of the four main methods for detecting ransomware infection. Zhao et al.~\cite{zhao} highlighted that the task of using entropy to discern compressed files from encrypted files is a challenging problem, as both types exhibit similar entropy values. Alekseev and Platonov~\cite{alekseev} modified the NIST Discrete Fourier Transform (Spectral) Tests, replacing the Fourier Transform with the entropy. They conducted experiments with four different encryption algorithms (3DES, AES256, XTEA, and BLOWFISH) and with 3 compression algorithms (UCL, LZMA, and APLIB), showing that encrypted files were successfully detected as potentially malicious and that compressed files were marked unencrypted. As expected, files that were first encrypted and then compressed were wrongly detected as unencrypted. However, the authors note that decompression algorithms are known and easy to apply, which makes it viable to decompress files and then reapply the modified spectral test.

Scaife et al.~\cite{scaife} use three primary indicators and two secondary indicators to identify whether a certain behaviour resembles a ransomware set of operations. Among the primary operations, the authors include the entropy, noting that encrypted files tend to have an entropy value close to the upper bound value returned by the Shannon entropy formula, calculated over an array of bytes. The authors observed that, across the 492 ransomware samples they used for their experiments, the main file types targeted by ransomware are common work-office documents, such as PDFs and DOCXs. This highlights that the fundamental strategy of ransomware is to encrypt files important enough for the victims to motivate them to pay a ransom. Therefore, some research investigated countermeasures tailored to productivity-related files.

Jung and Won~\cite{jung} proposed a context-aware framework that computes the entropy of header and footer of PDF files suspected of being infected with ransomware. Then, they compare the result with the expected entropy of PDF files, under the assumption that when ransomware encrypt entire files, it disrupts considerably the entropy value of such files. Although their experiments were limited to PDFs, it is worth noting that the approach can be extended to different file types.

Lee et al.~\cite{lee} addressed the problem of ransomware-infected files transferred to backup cloud services, which defeats the purpose of creating backups. In their work, the authors compare the entropy of a file against its backup version, before uploading it to the cloud. In case that the new version of the file is found to be infected, the cloud service transfers back to the local machine the last trustworthy backup version. The framework includes a machine learning module that analyses a number of different entropy values (e.g., the entropy measured with the collision test and the entropy measured with the Markov test) and detects whether an infection occurred or not. 

Hsu et al.~\cite{hsu2021} proposed a machine learning approach based on two different SVM models: SVM Linear and VSM kernel Trick (Poly). Using different features, including file types, multiple entropy, file compression ratio, and 0:1, the authors achieved detection rates of 85.17\% and 92\% for SVM linear and SVM kernel trick, respectively.

Some researchers draw attention to the limitations of using entropy values for performing ransomware detection. McIntosh et al.~\cite{mcintosh19} highlighted that many entropy-based mitigations are prone to spoofing attacks, where ransomware artificially alters the entropy of the encrypted files. Spoofing allows attackers to avoid detection by keeping the entropy below the warning threshold and, in turn, perform a successful ransomware infection. 
In their paper, the authors show that a simple application of a Base64-Encoding function to various encrypted files reduces their entropy from a value close to 8.0 (i.e., the upper threshold) to a lower value of 6.0. This result replicated consistently across 10 different file types, including PDFs, DOCXs, and CSVs. Moreover, the authors theorise an alternative kind of ransomware attack, struck by encrypting only portions of victims' files, in such a way that file disruption is maximised and encryption is minimised. This approach allows for keeping the entropy of the encrypted files low and, at the same time, to prevent information availability.

Pont et al.~\cite{pont} warn on the risk of using entropy as a sole indicator for ransomware infection. The authors used a dataset of over 80000 mixed files, such as images, compressed data, and encrypted data, and calculated the entropy values for each file type. Then, they compared the values against the thresholds used in the literature, for deciding whether a modification has been caused by ransomware or not. The authors experiments produced high values of False Rate Positives (FPRs), up to 92.80\%, leading to the conclusion that entropy alone (or other statistical approaches) might not suffice.

As aforementioned, calculating the entropy over thousands of files to detect ongoing ransomware infections is time-consuming and resource-consuming. Davies et al.~\cite{davies} developed a technique, named Differential Area Analysis (DAA), that focuses only on analysing file headers, instead of computing the entropy over entire files. As a result, not only their approach is faster than many approaches in the state-of-the-art, but it is also able to differentiate between compressed, legitimate encrypted, and ransomware encrypted files. Their approach shows that entropy can still be a valuable metric for ransomware detection, if used differently than simply computing it over an entire file. Indeed, their experiments consistently achieve accuracy results over 98\%, using a large dataset composed of more than 80000 files.

\section{Threat Model}\label{sec:threat-model}
This paper is split in two parts. In the first part, we propose attacks on the DAA defence, which involves modifying the behaviour of a traditional ransomware. In the second part, we propose a set of countermeasures against these modifications. In this section, we provide a brief threat model with assumptions and capabilities, both for the defender and the attacker.


Both DAA and the countermeasures proposed in this paper, try to detect ransomware during the encryption phase, when the mass-encryption of the victim's files has already started. This makes DAA and our countermeasures last-resort defences, to lean on when ransomware have escaped every other upstream defence in place. The defending goal is to acknowledge the infection and minimise as much as possible the damage, by sending a warning to the sysadmin and shutting down the infected machine. In this scenario, the defender:
\begin{itemize}
    \item Implements a real-time monitor that, after detecting a suspicious workload in terms of hard-disk I\textbackslash O events, invokes DAA or our countermeasures;
    \item Uses DAA, or our countermeasures, to analyse the files currently written and try to identify whether the file has been encrypted by ransomware;
    \item Receives a warning from the real-time monitor, in case an infection is detected, before the machine shuts down to preserve the remaining files.
\end{itemize}
This defence does not prevent ransomware from encrypting some files in the system, but it would shield the majority of the remaining files from infection. In turn, this would give time to the sysadmin to later analyse the hard-disk and recover the unaffected data.

With this defending scenario in mind, the attacker aims to evade the real-time monitor and either DAA or our countermeasures. We assume that the attacker:
\begin{itemize}
    \item Can modify the ransomware mass-encryption phase before deploying the attack, manipulating the entropy of the encrypted files in any way they deem necessary;
    \item Can manipulate in any way the file extensions;
    \item Cannot tamper with the real-time monitor. Every file modified by ransomware is necessarily analysed.
\end{itemize}

\section{Attacks on DAA}\label{sec:attacks}
DAA is based on the intuition that an encrypted file exhibits high entropy and a uniform distribution throughout the file. Based on this assumption, the algorithm limits the entropy analysis to a small portion of it. In particular, DAA focuses on the headers, which contain structural information for interpreting the rest of the files, hence exhibit low entropy values.
While DAA proved to be efficient and effective against current ransomware in the wild, the sole analysis of a file header makes it possible for new ransomware to manipulate the headers to bypass detection. 
Considering that the header is the only part of the file analysed by DAA, an attacker could manipulate the header to make its entropy indistinguishable from that of an unencrypted header (as described in \Cref{sec:threat-model}. In turn, this would entirely bypass the DAA defence.
This goal can be achieved in several ways. In this paper, we explore three alternatives:
\begin{itemize}
    \item Prepending to the file header a block of 256 bytes with low entropy, such as a single letter repeated to fill the block;
    \item Filling up the 256 bytes header with 32 repetitions of the first 8 bytes;
    \item Inserting low-entropy entries in random or fixed positions of the header, for example common words or repetitions of the alphabet.    
\end{itemize}
In this section, we show that the first two modifications decrease effectively the detection accuracy of DAA, while the last one is less effective.

\subsection{Attack Dataset}\label{sec:dataset}
To test the efficacy of our attacks, we start from the dataset created by Davies et al.~\cite{davies} which contains around 8000 files and includes:
\begin{itemize}
    \item Common files: DOC, DOCX, JPG, PDF, PPTX, XLS, XLSX, 7ZIP (with three different types of compressions), CSS, DLL, GIF, MKV, MP3, MP4, PNG, RAR, TAR and XML.
    \item Files encrypted by the following ransomware: Maze, Dharma, NetWalker, NotPetya, Phobos, Ryuk, Sodinokibi, BadRabbit and WannaCry.
\end{itemize}
To create the encrypted files, the authors took the common files in their dataset (in particular, DOC, DOCX, JPG, PDF, PPTX, XLS, and XLSX) and applied the different ransomware encryption.

To simulate the results that ransomware would obtain by applying the three attacks that we propose, we enriched Davies et al. dataset by adding modified encrypted files. First, we took 100 files from each file type in the dataset provided by Davies et al., for a total of 2900 files (2000 common files and 900 encrypted files). Then, to test the three attacks, we modified the header of every ransomware encrypted file as follows:
\begin{itemize}
    \item Files with suffix ``low-H'', prepending a low-entropy 256 bytes block (i.e., a repetition of the character 'a'). With 1 byte per character, we obtain 256 repeated characters, making the entropy of the header extremely low;
    \item Files with suffix ``rep-bytes'', repeating the first 8 bytes for 32 times. In this second case, the number of repetitions is smaller than the first approach, therefore the entropy of the header is not as low as with the first approach;
    \item Files with suffix ``com-seq'', substituting the first 256 bytes of the file with a random succession of various low-entropy sequences. We used the uppercase alphabet, the lowercase alphabet, some common words, numbers between 0 and 22, and a repetition of the number zero. This third attack is still capable of reducing the entropy, but not as much as the previous two attacks, due to its fewer character repetitions.
\end{itemize}

In conclusion, our dataset is composed of 5600 files, split in 2000 common files and 3600 encrypted by ransomware.

\subsection{Results of our Attacks}\label{par:daa-orgres}\label{par:daa-mixres}
\begin{figure*}
     \centering
     \begin{subfigure}[t]{\columnwidth}
         \centering
         \includegraphics[trim={0 0 0 0.5cm},clip,width=\textwidth]{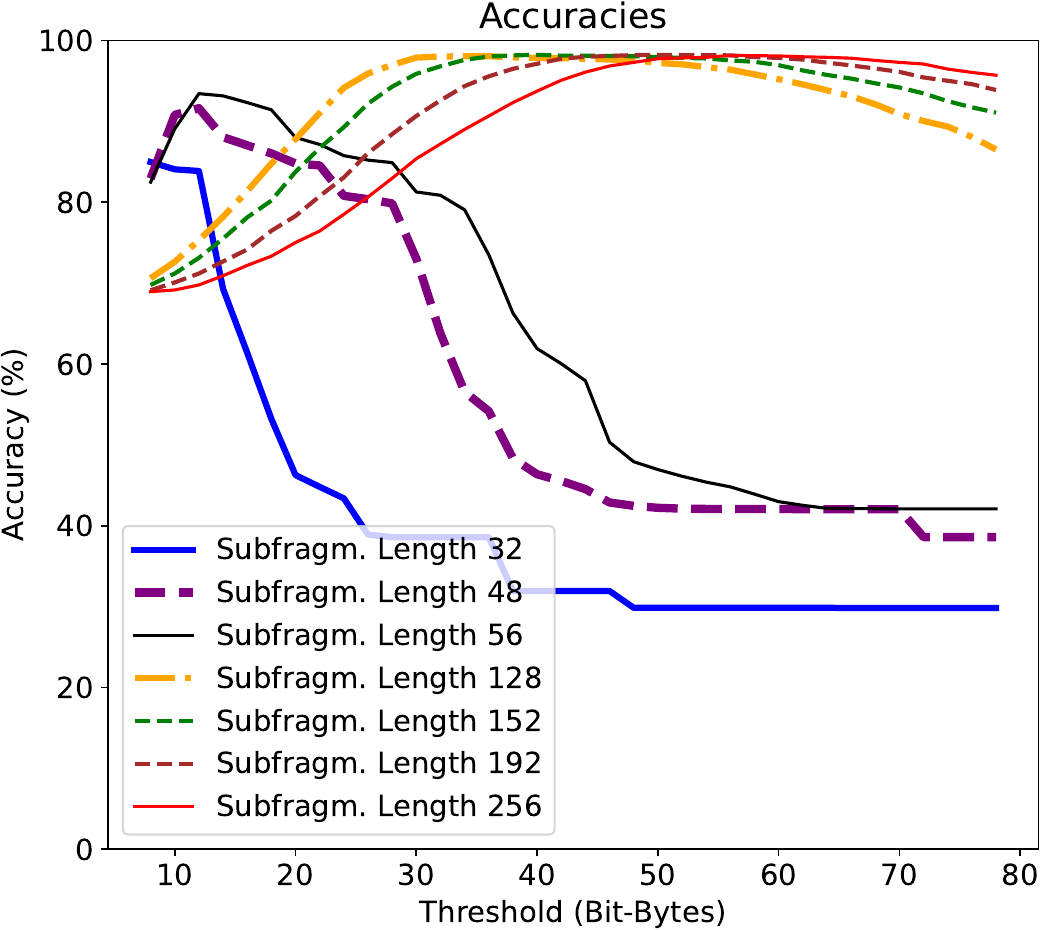}
         \caption{Original dataset.}
         \label{img:daa-org}
     \end{subfigure}
     \hfill
     \begin{subfigure}[t]{\columnwidth}
         \centering
         \includegraphics[trim={0 0 0 0.5cm},clip,width=\textwidth]{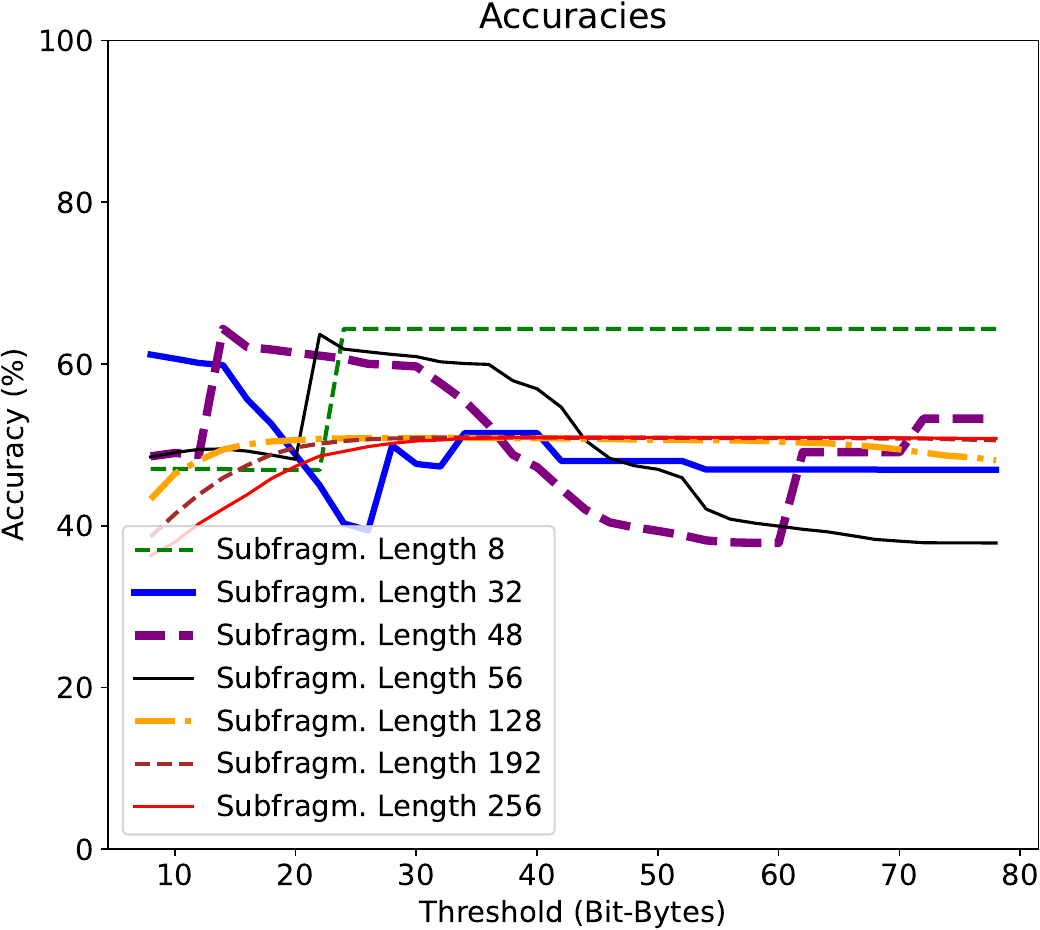}
         \caption{Attack dataset.}
         \label{img:daa-forg}
     \end{subfigure}
        \caption{DAA classification accuracy results, applied to the original dataset and the attack dataset that we created.}
        \label{fig:daa-before-after}
\end{figure*}

As shown in the first column of \Cref{table:daa-attacked}, our tests confirm Davies et al.~\cite{davies} conclusions and show that DAA on the original dataset achieves optimal results for each metric (i.e., accuracy, precision, recall, and F1 score). In \Cref{img:daa-org}, which highlights the overall accuracy for each header length and threshold used, it is easy to notice that the larger the header length, the higher the accuracy achieved. This is consistent with the mode of operation of DAA, and clearly confirms that DAA still works as expected on the original dataset we have utilised, albeit in its reduced form.

\begin{table}[t]
\centering
\small
\renewcommand\arraystretch{1.5}
\caption{Best accuracy obtained with DAA against the original dataset and the attack dataset.}
\label{table:daa-attacked}
\begin{tabular}{lll}
 & DAA on Original Data & DAA on Attack Data \\ 
\toprule
Length & 152 & 48 \\
Threshold & 40 & 14 \\
Accuracy (\%) & \textbf{98.24} & 64.34 \\
Precision (\%) & \textbf{99.30} & 92.30 \\
Recall (\%) & \textbf{95.00} & 48.64 \\
F1 (\%) & \textbf{97.10} & 63.71 \\
\bottomrule
\end{tabular}
\end{table}

However, when DAA is applied to the attack dataset that we created as explained in \Cref{sec:attacks}, the results change. As shown in the second column of \Cref{table:daa-attacked}, the metrics worsen across the board. In the best case scenario, with a subfragment length of 48 and a threshold of 14, the accuracy and F1 score decrease by more than 30\% and the recall by almost 50\%. Although not as critically as the other metrics, the precision also decreases from 99.30\% to 92.30\%.

The good performance on the precision metrics highlights the fact that DAA is still able to keep a low amount of false positives (i.e., most unencrypted files are correctly classified). However, the low recall indicates that DAA incorrectly classifies many encrypted files as unencrypted, with respect to the actual number of encrypted files. In particular, almost every file we forged was erroneously classified as unencrypted and escaped DAA detection. 
In summary, comparing the results in \Cref{img:daa-org} with the results shown in \Cref{img:daa-forg}, it looks clear that our attacks are effective in decreasing the performance of DAA. This is consistent with the mechanisms of our attacks, capable of reducing the entropy of ransomware-encrypted files to the entropy levels of legitimate files. In other words, our attacks invalidate the underlying assumption of DAA that the entropy of ransomware-encrypted files is distinguishable from other categories of files. 


Although \Cref{table:daa-attacked} shows the result of the entire dataset, containing files of all our three attacks, it is interesting to analyse more in depth how effective each attack is. In the interest of time and space, we do not include the detailed data breakdown, but we provide the following insights, to shed some light.
Our preliminary analysis on DAA performance against each of our attacks, show that ransomware adopting the low-H and rep-bytes attack strategies manages to clearly decrease the detection accuracy of DAA. In contrast, the com-seq attack seems to be ineffective, with DAA achieving almost 100\% detection accuracy. 

\begin{figure}[t]
\centering
\includegraphics[width=\linewidth]{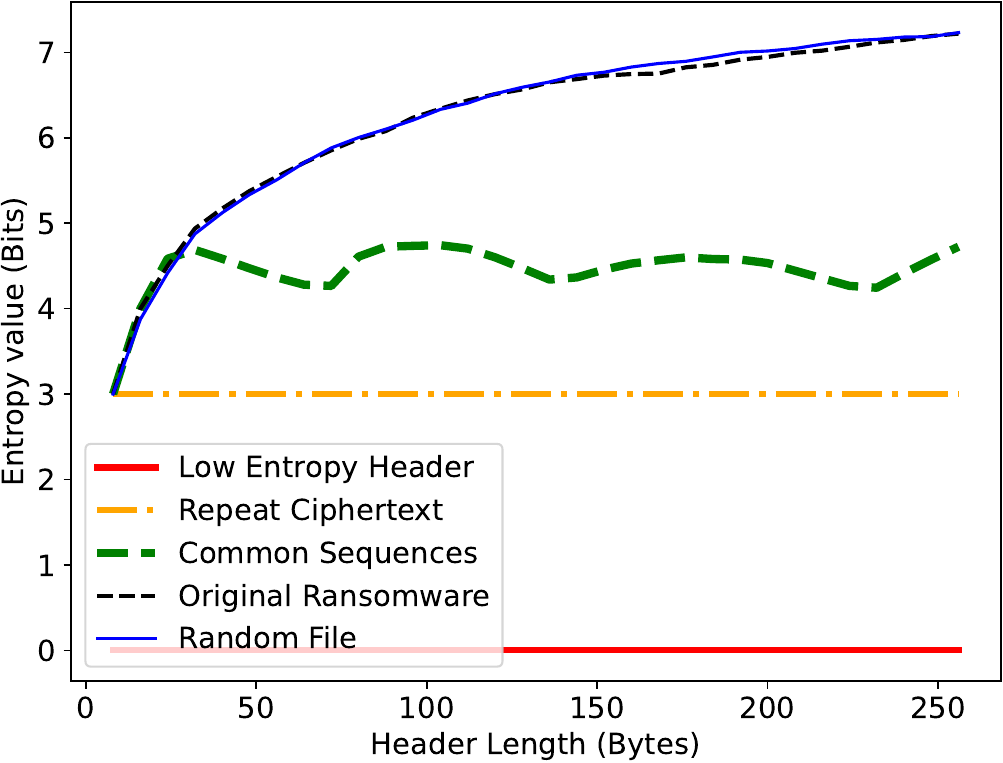}
\caption{Example entropy values for a random file, a ransomware-encrypted file, and three ransomware-encrypted files with low-H modification, rep-bytes modification, and com-seq modification, respectively.}
\label{img:com-seq-1}
\end{figure}

This result can be explained by analysing the specifics of how the entropy of the headers change, for each of the three attacks. In \Cref{img:com-seq-1}, we show an example of how each attack affects (i.e., decreases) the header entropy, compared to the result of the original ransomware attack. The attack com-seq (common sequences) is the one that decreases the entropy the least, even though the differential area is still considerably large if we consider the entire header. This indicates that com-seq can be an effective attack against DAA, if DAA uses the original header length of 152 bytes. However, if DAA uses a fragment length of 48 bytes, optimised for the attacks as shown in \Cref{table:res-mix}, then the accuracy drops to almost 0\%. The reason is clear when looking at \Cref{img:com-seq-1}: the differential area calculated between the green curve and the blue curve is minimal with a 48 bytes header length, hence DAA is able to hit an accuracy of 100\%. 

These results are in line with our expectations, in light of the effects that our three attacks have on the entropy, as we discussed in \Cref{sec:dataset} and exemplified in \Cref{img:com-seq-1}. From this analysis, we derive that to make our attacks as effective as possible, it is not enough to lower the average entropy of the header, but it is critical to lower the entropy of the first 50 bytes. In the next section, we describe which countermeasures we propose for mitigating these attacks.

\section{Countermeasures Against our Attacks}\label{sec:countermeasures}
The attacks mentioned in \Cref{sec:attacks} are made possible by the fact that DAA focuses on analysing only a predefined and limited part of a file. Even though effective, this makes the approach predictable and easy to circumvent. Therefore, it is necessary to strengthen DAA in such a way that it can resist to malicious manipulation of file headers. In this section, after describing the necessary terms, we propose three approaches for improving DAA robustness.

\subsection{Parameters used for the analysis}
Before describing the mitigation we developed, it is necessary to define the three different parameters that we use in our algorithms: \textit{threshold, distance, and subfragment length}. As shown later in \Cref{sec:count_results}, in our experiments, we investigate which combination of the three parameters yields the best results in terms of identification accuracy.

\paragraph*{Threshold} As explained in \Cref{sec:relatedworks}, when performing the differential area analysis, it is necessary to establish the range within the value is considered too close to the curve of a random file, which signals that the file is encrypted.

\paragraph*{Distance} In regular unencrypted files, fragments positioned far away from the header have a relatively high chance to exhibit high entropy, since structural information is stored between file portions close to the header.

Since our mitigations analyse the header fragment and one (or more) additional fragments, it is necessary to take into account the higher entropy (hence, the smaller differential area) of the random-position fragment. This is necessary to prevent a skewed behaviour in our algorithms, as shown later in \Cref{alg:mit2} and \Cref{alg:mitN}. To do so, we introduce a parameter named \textit{distance}, which lowers the value of the differential area related to the header fragment when comparing it to the differential area of the random-position fragment(s). This has the effect of increasing the priority of the header entropy information over the randomly picked fragment, which is especially important when analysing unencrypted files. By tuning the distance parameter, we lower the false positives occurrences, hence we reduce the chance to wrongly classify regular files as ransomware-encrypted files.

\paragraph*{Subfragment length} DAA analyses 256 bytes in incremental subfragments that grow from 8 up to 256 bytes, 8 bytes at a time. At each step, the detection accuracy changes, and in some cases analysing fewer steps than 32 can yield to better results. Therefore, it is useful to keep track of the accuracy achieved at each incremental step and correlate it to the analysed length of the subfragment.

\subsection{Mitigations Overview}\label{sec-mit-ov}
As mentioned at the beginning of this section, DAA~\cite{davies} focuses on analysing the first 256 bytes of a file, making it easy for attackers to modify their ransomware and counteract the detection. Therefore, it is necessary to reduce the predictability of DAA. In this paper, we propose to do this by analysing additional fragments in the file at random positions, and we investigate the ideal number of fragments to use. In particular, we investigate the results of adding one, two, and three additional fragments to the baseline DAA method:
\begin{itemize}
    \item 2-Fragments (2F) analyses the header and a random fragment of the file. The length in bytes of both fragments is a multiple of 8, up to 256 bytes. The algorithm calculates the differential area for the fragment and the header, applying the \textit{distance} parameter to the area of the header as previously discussed. Last, it selects the lower differential area (i.e., the fragment that exhibits the greater entropy).
    \item 3-Fragments (3F) works similarly to 2F, with the addition of a second random fragment. First, the algorithm calculates the differential area of the two random fragments and computes the average. Then, as done for 2F, it applies the \textit{distance} parameter to the differential area of the header, compares it to the average area of the two random fragments, and selects the lower differential area.
    \item 4-Fragments (4F) runs as 3F, with the only difference that the random fragments are three, hence it calculates the average differential area over three differential areas.
\end{itemize}

The outcome of the differential area calculation is an area unit measured in Bit-Bytes, representing the fragment length in bytes and the related byte entropy in bits. As aforementioned, the lower the calculated differential area, the higher the likelihood that the file contains encrypted data.

To compute the total differential area for each subfragment length, we use the Composite Trapezoidal Rule described by Atkinson~\cite{atkinson}:
\begin{equation}\label{eq:trap-rule}
    I_{composite} = \frac{h}{2} \cdot [f(a)+2 \cdot \sum f(x+h_{i})+f(b)].
\end{equation}
Let us assume that we analyse a fragment of 256 bytes by incremental steps of 8 bytes. Then, $a = 8$ represents the length of the first subfragment to analyse, and $b = 256$ is the length of the last subfragment to analyse. The term $h = 8$ indicates the step value between each point, and $f$ is the function that outputs the byte entropy, given the subfragment as input.

\subsection{Mitigations Algorithms and Implementation}\label{sec:mits}
In this section, we describe in detail the algorithms that we developed for implementing our countermeasures with the related pseudocodes. We implemented the algorithms using Python 3 and ran the experiments on Jupiter Notebook. The code is open-source and is available on GitHub~\cite{github}. Before we delve in the details of our three countermeasures, we would like to highlight that we always select the header as one of the fragments. As observed by Davies et al.~\cite{davies}, the header carries lots of relevant information in the case of legitimate files. Selecting the header aligns our countermeasures to this observation, ensuring good performance not only against our attacks, but also on legitimate files and traditional ransomware-encrypted files.

\paragraph*{\textbf{2-Fragments (2F) Countermeasure}}\label{par:exp2F} As previously described, DAA analyses one fragment of 256 bytes. With 2F, we analyse two fragments, one starting at position 0 (i.e., the header) and one picked at a random byte position, within the range $[o, ..., length(file)-o]$. 
Let us analyse the algorithm for the 2F mitigation, shown in \Cref{alg:mit2}, assuming fixed threshold, distance, and fragment length.

First, we create the vector \textit{randvalues} containing \textit{o} random bytes, such that we have a reference fragment for performing the differential area analysis. For each subfragment $i \in [i \cdot 8]$, first the algorithm computes the entropy. Each entropy value is used to calculate the differential area for the header $areaH$ and for the fragment selected at a random position $areaR$. Then, given the distance parameter \textit{d}, the algorithm subtracts \textit{d} from $areaH$. If $areaR$ is between $areaH$ and $areaH - d$, $areaH$ is stored in $area$. Else, the algorithm stores in $area$ the minimum between $areaH$ and $areaR$.
As explained earlier in the description for the 2F approach, \textit{d} helps to increase the weight of the header on the final result, as we observed that in some occasions (such as with unencrypted files) it tends to be more reliable than the result obtained with a random fragment. In fact, without subtracting \textit{d} as shown in the for-loop at line 15 of \Cref{alg:mit2}, the algorithm would correspond to simply choosing the smaller differential area, hence the result that yields the greatest entropy.
Finally, the algorithm compares $area$ to the threshold $t$ and returns $True$ if the area is smaller than $t$, signalling the presence of ransomware. Otherwise, it returns $False$.

\begin{algorithm}[th]
\caption{2-Fragments Mitigation}
\label{alg:mit2}
\hspace*{\algorithmicindent} \textbf{Input}: List of files ($files$) to analyse, optimal distance ($d$), threshold ($t$) and fragment length ($o$)\\
\hspace*{\algorithmicindent} \textbf{Output}: True if the file is encrypted by ransomware, False otherwise
\begin{algorithmic}[1]
\State $randvalues[i] \gets randint(0,255)$ for i in [1, o]
\ForEach {$file$ in $files$}
    \State $buf \gets open(file)$
    \For{$len=8; len \leq o; len=len+8$}
        \State $h[len/8] \gets entropy(buf, 0, len)$
        \State $h ideal[len/8] \gets entropy(randvalues, 0, len)$
    \EndFor
    \State $areaH \gets diffarea(h ideal - h)$
    \State $start \gets randint \in[o, size(buf) - o -1]$
    \For{$len=8; len \leq o; len=len+8$}
        \State $h[len/8] \gets entropy(buf, start, len)$
        \State $h ideal[len/8] \gets entropy(randvalues, 0, len)$
    \EndFor
    \State $areaR \gets diffarea(h ideal - h)$
    \If{$areaH > areaR$ and $areaH- d < areaR$}
        \State $area \gets areaH$
    \Else
        \State $area \gets min(areaH, areaR)$
    \EndIf
    \If{$area \geq t$} 
        \State $return$ False
    \Else
        \State $return$ True
    \EndIf
\EndFor
\end{algorithmic}
\end{algorithm}

\paragraph*{\textbf{3-Fragments (3F) Countermeasure}}\label{par:exp3F} 
In this approach, instead of analysing the header and one additional fragment selected at a random point of the file, we analyse two fragments in addition to the header. \Cref{alg:mitN} shows the generic algorithm for an arbitrary number of $N-1$ additional fragments. For describing the 3F countermeasure, we instantiate the generic algorithm as three fragments in total.

As previously done for \Cref{alg:mit2}, we create the vector \textit{randvalues} containing \textit{o} random bytes. Then, we split the input file in three partitions of equal length $b$, as shown at line 7 of \Cref{alg:mitN}. After computing the differential area of the first fragment (the header) and storing it in the vector $areas$, the algorithm chooses the starting byte position $start$ of the next fragment to analyse. Position $start$ is randomly chosen within a range between $n \cdot b$ and $(n+1) \cdot b$, where $n \in [1,N]$. Then, the algorithm computes the differential area of the fragment and stores it in vector $areas$ (\Cref{alg:mitN}, line 11), as done with the differential area value of the header.

The main difference between 2F and 3F lies in how the differential areas are used. In 2F, the differential area of the header was directly compared to the differential area of the (only) additional fragment. In 3F, the differential area of the header fragment $areas[1]$ minus the distance \textit{d} (i.e., $areas[1] - d$) is compared with the average $avg$ of the differential areas of the additional two fragments ($areas[2]$ and $areas[3]$). Again, the smaller value is stored in the final variable $area$ and finally compared with the threshold $t$.

\begin{algorithm}[t]
\caption{N-Fragments Mitigation}
\label{alg:mitN}
\hspace*{\algorithmicindent} \textbf{Input}: List of files ($files$) to analyse, optimal distance ($d$), threshold ($t$), fragment length ($o$) and number of fragments to analyse ($N$)\\
\hspace*{\algorithmicindent} \textbf{Output}: True if the file is encrypted by ransomware, False otherwise
\begin{algorithmic}[1]
\State $randvalues[i] \gets randint(0,255)$ for i in [1, o]
\ForEach {$file$ in $files$}
    \State $buf \gets open(file)$
    \State $b \gets size(buf) / N$
    \State $start \gets 0$
    \For{$n=1; n \leq N; n=n+1$}
        \For{$l=8; l \leq o; l=l+8$}
            \State $h[l/8] \gets entropy(buf, start, l)$
            \State $h ideal[l/8] \gets entropy(randvalues, start, l)$
        \EndFor
        \State $areas[n] \gets diffarea(h ideal - h)$
        \State $start \gets randint \in[n*b, (n+1)*b]$
    \EndFor
    \State $avg \gets mean(areas[2:])$
    \If{$areas[1] - d < avg$}
        \State $area \gets areas[1]$
    \Else
        \State $area \gets min(avg, areas[1])$
    \EndIf
    \If{$area \geq t$} 
        \State $return$ False
    \Else
        \State $return$ True
    \EndIf
\EndFor
\end{algorithmic}
\end{algorithm}

\begin{figure}[t]
\centering
\includegraphics[width=\linewidth]{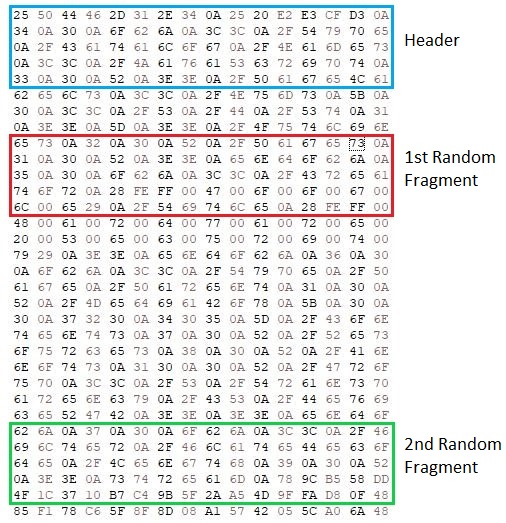}
\caption{Some byte fragments of length 80 inside a normal PDF file that may be chosen by the 3F algorithm: the header fragment (blue one), and the two random fragments (red and green ones).}
\label{img:distance}
\end{figure}

Let us provide an example on how the fragments are chosen, assuming the use of the 3F algorithm on a PDF file, as shown in \Cref{img:distance}. The three coloured boxes in \Cref{img:distance} highlight the three fragments chosen by 3F for analysing the file and deciding whether it has been encrypted or not. Let us also assume that the differential area calculated for the header is $D_{H} = 71$, for the first random fragment is $D_{F1} = 32$, and for the second random fragment is $D_{F2} = 54$. In this case, the predefined value for the threshold is $t = 46$ and for the distance is $d = 35$. The goal is to select one of the three differential areas and compare it with the threshold to determine if the file is suspicious or not. The selection process follows these steps:

\begin{enumerate}
    \item Compute the differential areas of the random fragments and calculate the average $avg$. In our example, $(D_{F1} + D_{F2}) / 2 = (32 + 54) / 2 = 43$;
    \item Subtract distance $d$ from the differential area of the header $D_{H}$,  $D_{H}-d=36$. Then, compare the latter result with $avg=43$;
    \item $D_{H} - dist$ is smaller than $avg$ (i.e., $36 < 43$). Therefore, the differential area selected is $D_{H} = 71$.
\end{enumerate}

Last, to classify the file as encrypted or unencrypted, the algorithm compares the selected area with the threshold. Referring to our example, $D_{H} = 71$, $t = 46$. Since $D_{H} > t$, the file is classified as unencrypted.
\\
The example showcases the importance of introducing the distance value for preventing the occurrence of false positives. In this example, without the use of a $d$ parameter, the minimum value chosen would have been $avg = 43$ which is smaller than the threshold $t=46$, leading to erroneously classify the normal PDF file as an encrypted file. This scenario is not unusual, as in normal files the entropy tends to grow larger with the increasing distance between a random fragment and a header fragment. Therefore, the distance parameter helps to give a higher weight to the header fragment as it is generally more reliable for classification purposes, especially with unencrypted files.

\paragraph*{\textbf{4-Fragments (4F) Countermeasure}}\label{par:exp4F} 
Similarly to the 3F countermeasure explained in \Cref{par:exp3F}, the generic algorithm for N-Fragments \Cref{alg:mitN} shows the pseudocode for our third countermeasure. The steps are the same for both countermeasure, with the only difference that, while 3F uses the header and two additional random fragments, 4F analyses the header and three random fragments.

\section{Countermeasures Results}\label{sec:count_results}
In this section, first we assess the efficacy of the three countermeasures we proposed (2F, 3F, and 4F) against the attacks described in \Cref{sec:attacks}. We applied the three approaches to the attack dataset that contains the files modified according to our attacks, mixed with the original files of Davies et al. dataset. Then, we analyse the performance of our countermeasures on the original dataset, to ensure that they would not render DAA ineffective against traditional ransomware. Then we analyse the time analysis performance of the original DAA against DAA modified with our countermeasures, to assess the overhead. Last, we put our algorithms under stress test, to evaluate how much an attacker would need to manipulate a ransomware-encryption result, to render our countermeasures void.

\begin{table}[t]
\caption{Best results of 2F, 3F and 4F countermeasures on the attack dataset.}
\label{table:res-mix}
\centering
\normalsize
\renewcommand\arraystretch{1.5}
\begin{tabular}{llll}
Data & 2F & 3F & 4F \\ 
\toprule
Length & 48 & 48 & 48\\
Threshold & 12 & 10 & 14\\
Distance & 54 & 48 & 56\\
Accuracy (\%)  & \textbf{92.78} & 91.75 & 92.53\\
Precision (\%) & \textbf{94.51} & 93.53 & 94.22\\
Recall (\%) & \textbf{94.25} & 93.64 & 94.17\\
F1 (\%)  & \textbf{94.38} & 93.59 & 94.19\\
\bottomrule
\end{tabular}
\end{table}

\subsection{Mitigations Against Attack Dataset}\label{sec:mit-attack}
In \Cref{table:res-mix}, we show the results obtained when applying our approaches to the attack dataset that we created. First, comparing the best result of every countermeasure (2F, 3F, and 4F) with the best result obtained by DAA and showed in \Cref{table:daa-attacked}, the accuracy increases by about 27\%. Indeed, while DAA obtained a best accuracy result of 64\%, our three approaches achieve an accuracy of over 91\%, with 2F reaching the best accuracy of 92.78\%. Previously, in \Cref{table:daa-attacked}, we showed that recall and F1 score for DAA decreased noticeably. On the contrary, every countermeasure we proposed achieve results over 93\% for both metrics, against a 48.64\% recall and a 63.71\% F1 obtained by DAA. Last, DAA yielded a precision of 92.30\% against the attack dataset and our countermeasures show an equally good performance with marginal improvements.
It is worth noting that the best results for every approach are obtained with a subfragment length of 48 bytes, and that the values for the threshold and the distance slightly change across countermeasures.

\begin{figure}[tpb]
\captionsetup[subfigure]{aboveskip=2pt,belowskip=4pt}
     \begin{subfigure}[t]{\linewidth}
         \includegraphics[width=\textwidth]{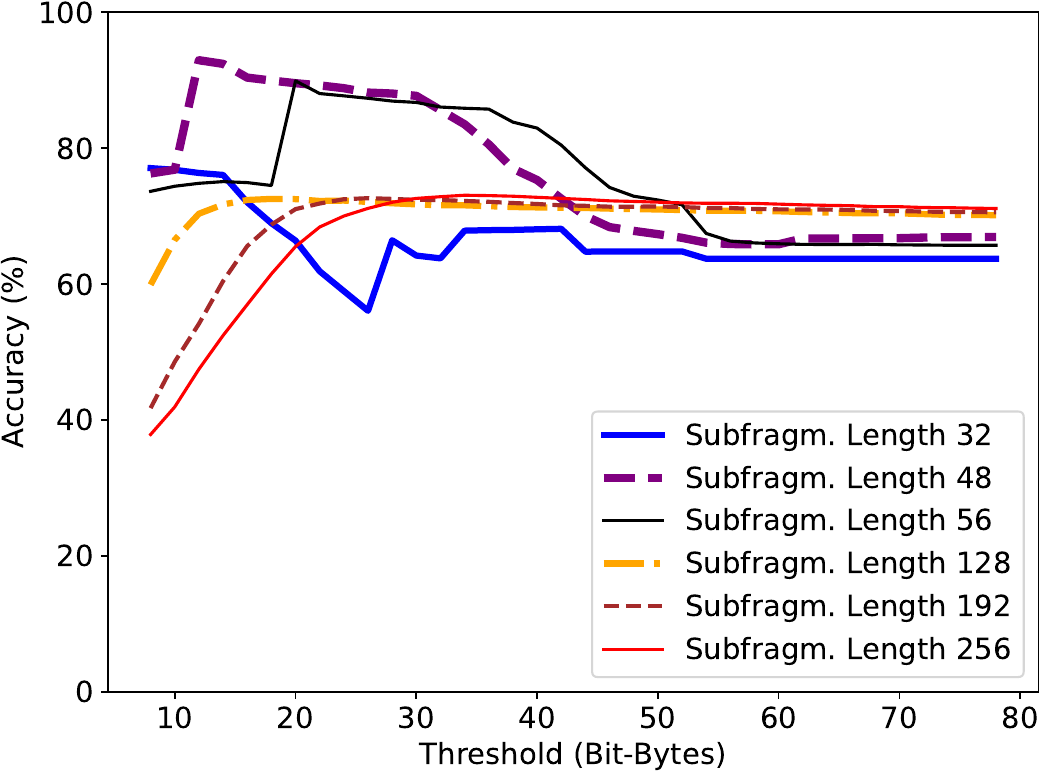}
         \caption{2F countermeasure}
         \label{img:2f-forg}
     \end{subfigure}
     \begin{subfigure}[t]{\linewidth}
         \includegraphics[width=\textwidth]{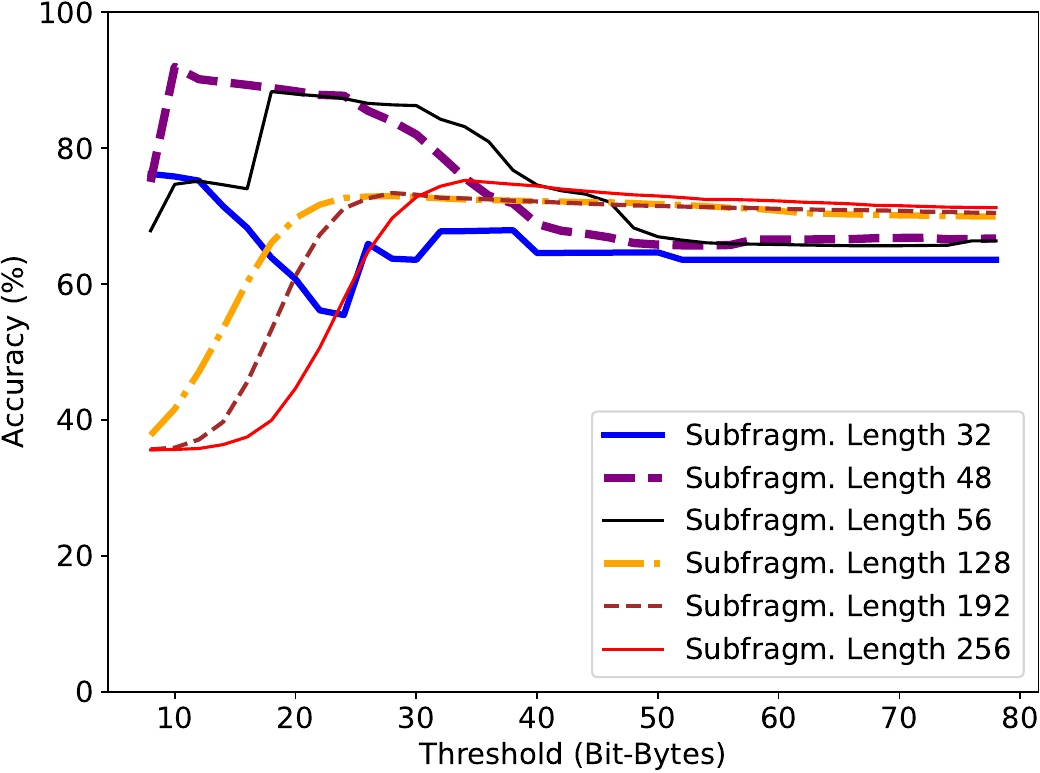}
         \caption{3F countermeasure}
         \label{img:3f-forg}
     \end{subfigure}
     \begin{subfigure}[t]{\linewidth}
         \includegraphics[width=\textwidth]{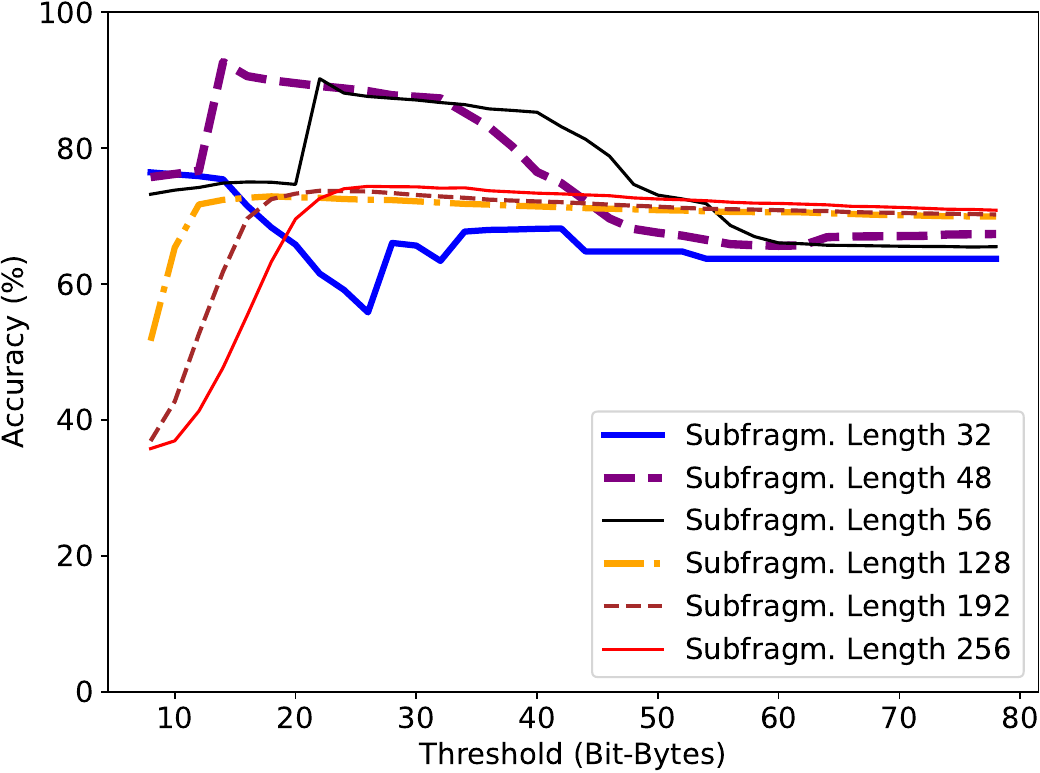}
         \caption{4F countermeasure}
         \label{img:4f-forg}
     \end{subfigure}
        \caption{Overall classification accuracy for each countermeasure on the attack dataset. Best distance parameter fixed.}
        \label{fig:acc_distance_fixed}
\end{figure}

The accuracy results for 2F, 3F, and 4F for different values of threshold and subfragment length are illustrated in \Cref{img:2f-forg}, \Cref{img:3f-forg}, and \Cref{img:4f-forg}, respectively. The three graphs are similar since they all show good best results when using subfragment lengths that equal 48 and 56. With larger thresholds, the accuracy tends to decrease to a point where subfragments larger than 56 bytes yield slightly better results.

\begin{figure}[tpb]
\captionsetup[subfigure]{aboveskip=2pt,belowskip=4pt}
     \begin{subfigure}[t]{\linewidth}
         \centering
         \includegraphics[width=\textwidth]{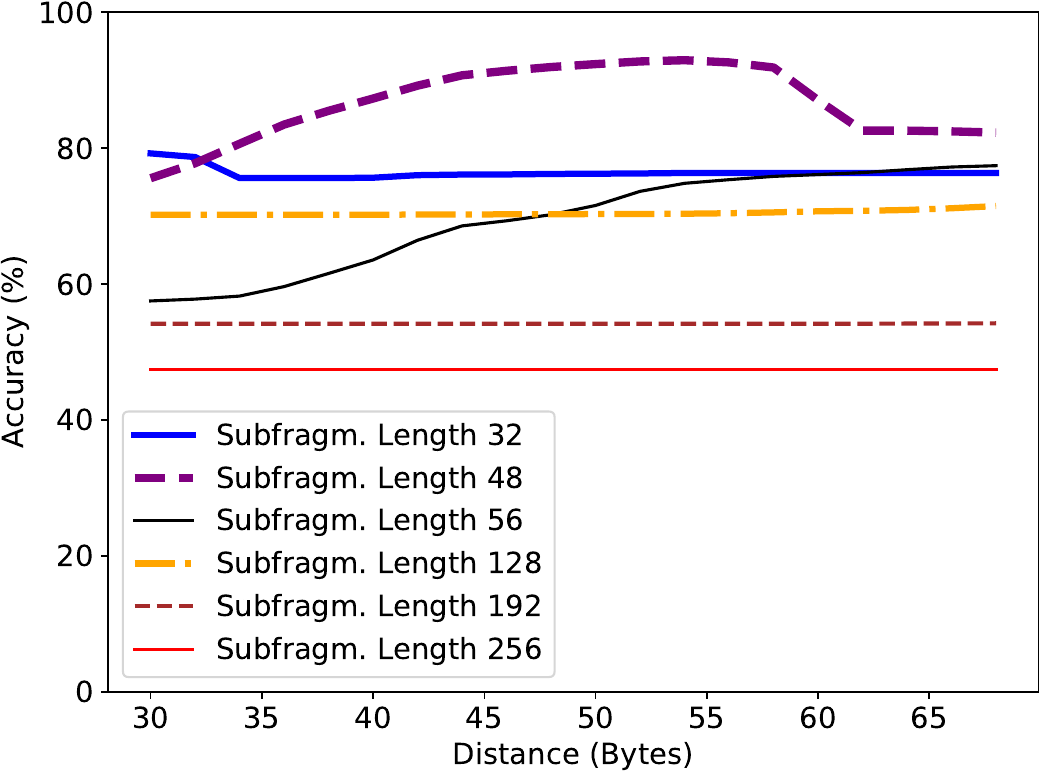}
         \caption{2F countermeasure}
         \label{img:2f-forg-dist}
     \end{subfigure}
     \begin{subfigure}[t]{\linewidth}
         \centering
         \includegraphics[width=\textwidth]{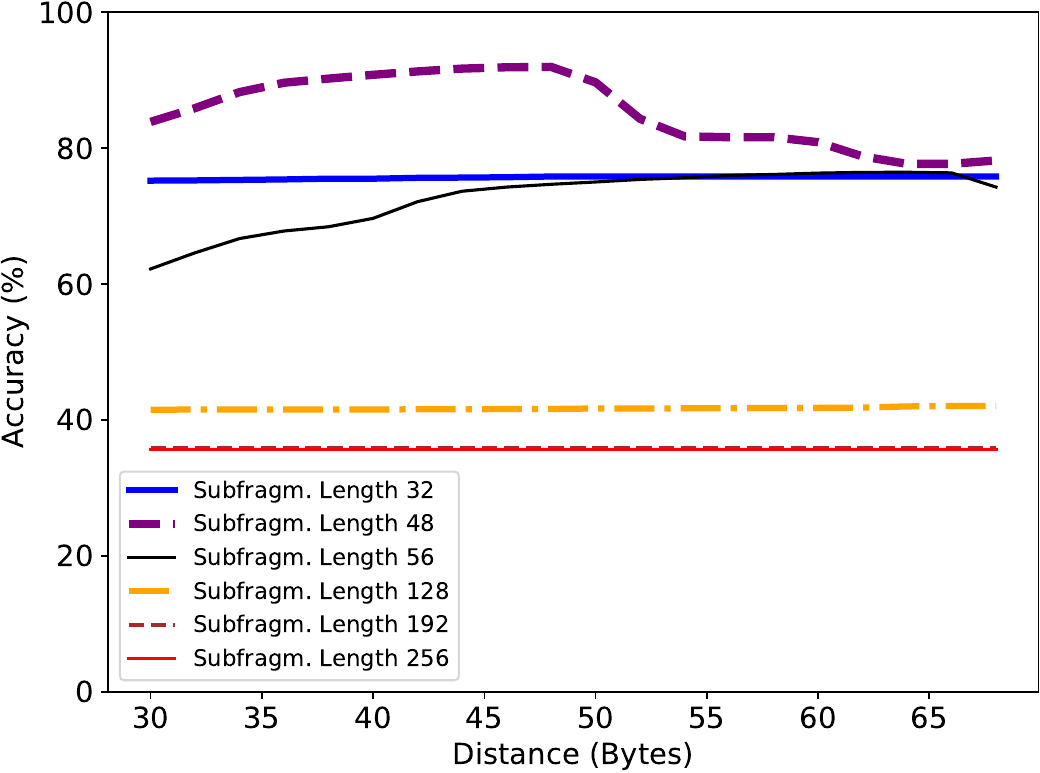}
         \caption{3F countermeasure}
         \label{img:3f-forg-dist}
     \end{subfigure}  
     \begin{subfigure}[t]{\linewidth}
         \centering
         \includegraphics[width=\textwidth]{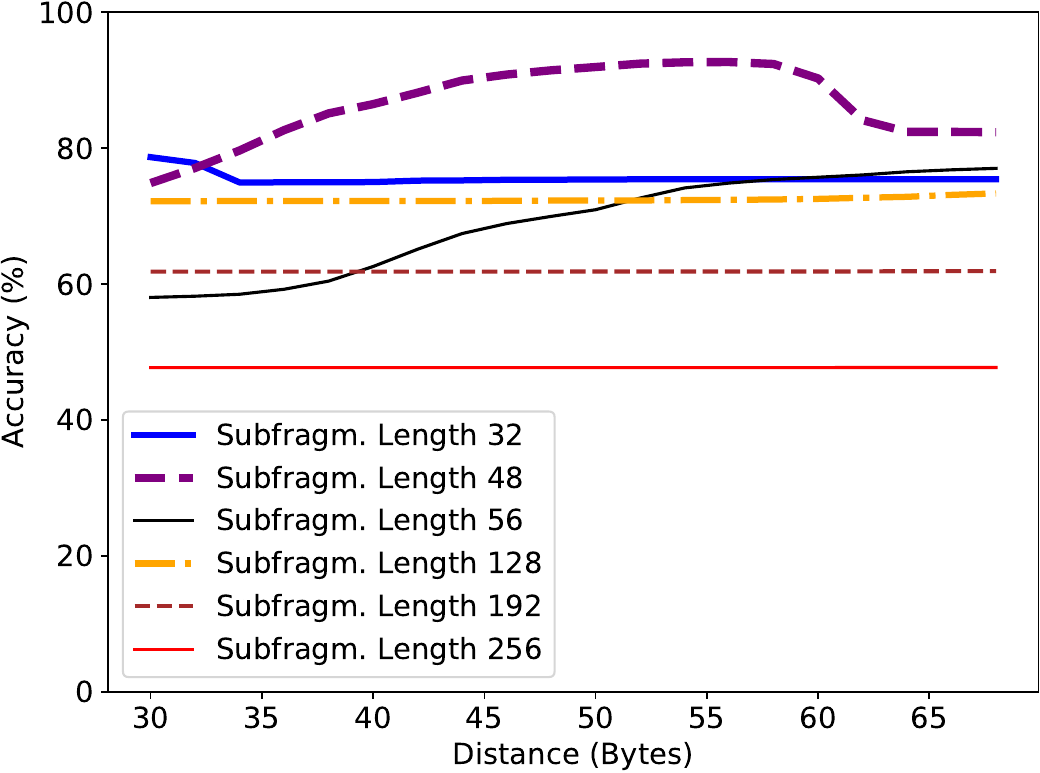}
         \caption{4F countermeasure}
         \label{img:4f-forg-dist}
     \end{subfigure}
        \caption{Overall classification accuracy for each countermeasure on the attack dataset. Best threshold parameter fixed.}
        \label{fig:acc_threshold_fixed}
\end{figure}

Similarly, the accuracy results of our mitigations at varying values of distance and subfragment length are illustrated in \Cref{img:2f-forg-dist}, \Cref{img:3f-forg-dist}, and \Cref{img:4f-forg-dist}, respectively. We can notice that, for all of these plots, the accuracy remains stable for the majority of the subfragment values. Each of the three graphs shows a similar behaviour with respect to the varying distance value. Before reaching the best value, as the distance increases, the accuracy increases marginally but constantly. However, once the best value is reached, the performance shows a sudden hit and the accuracy graph line exhibits a sudden drop.

\begin{figure*}[tpb]
\captionsetup[subfigure]{aboveskip=2pt,belowskip=4pt}
     \begin{subfigure}[t]{\columnwidth}
         \includegraphics[width=\textwidth]{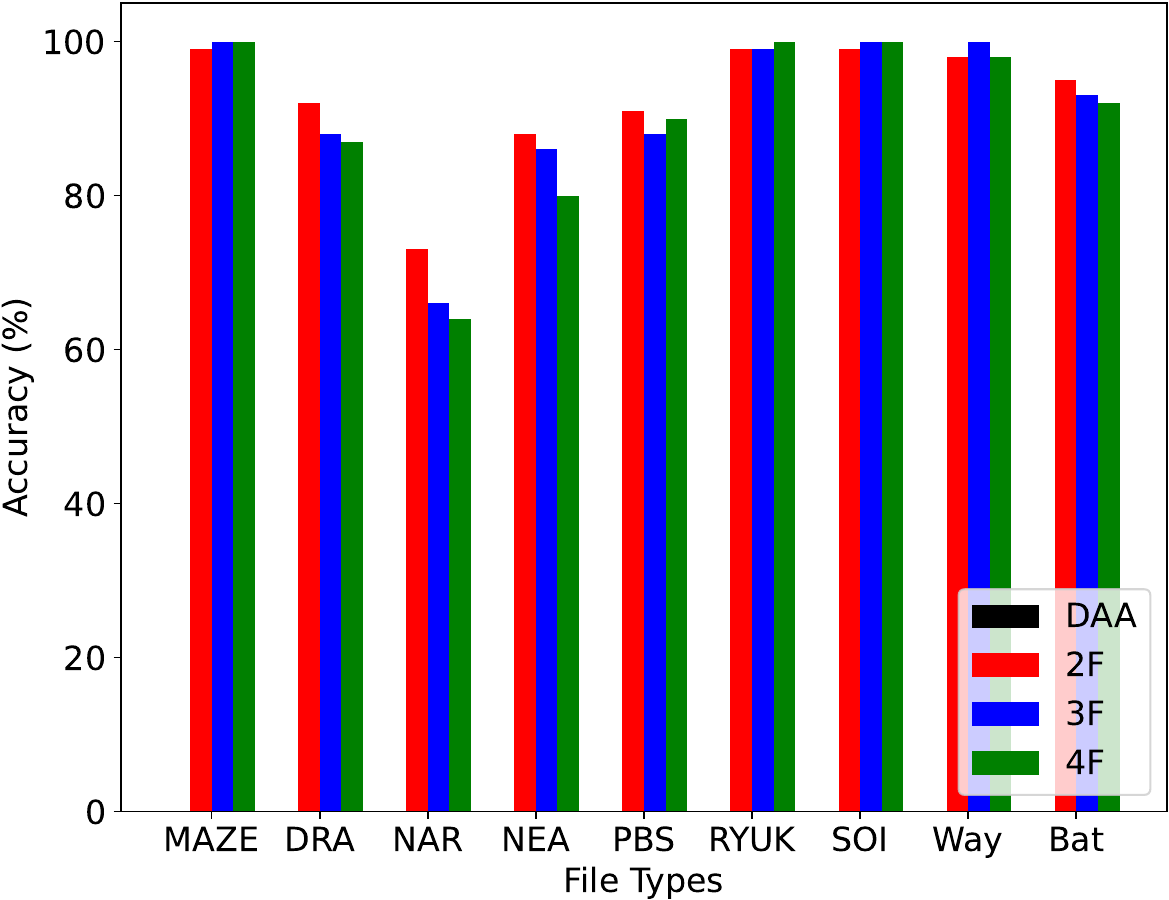}
         \caption{Low-H forged files}
         \label{img:low-H}
     \end{subfigure}
     \hfill
     \begin{subfigure}[t]{\columnwidth}
         \includegraphics[width=\textwidth]{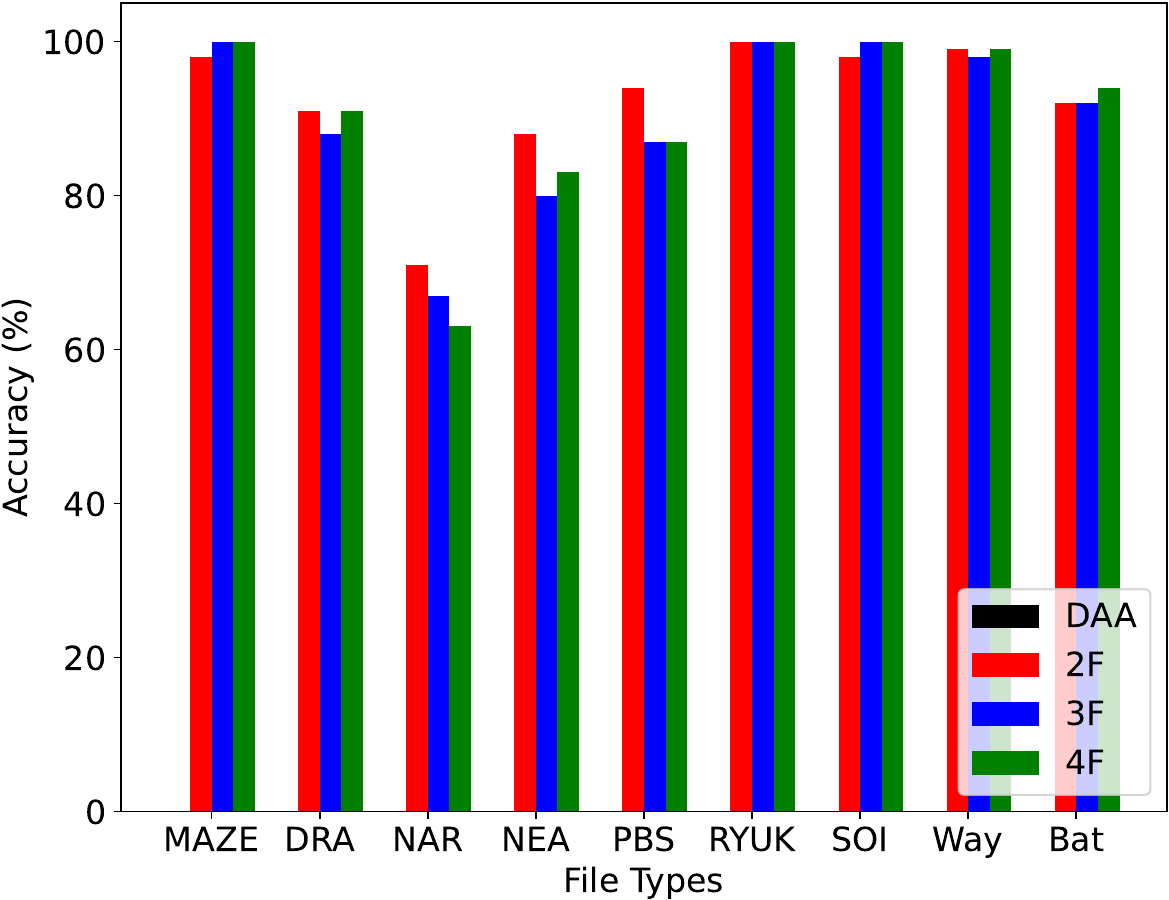}
         \caption{Rep-bytes forged files}
         \label{img:rep-bytes}
     \end{subfigure}
        \caption{Classification accuracy of DAA and our three mitigations, applied to files forged using the low entropy header (low-H) attack and the repeated bytes (rep-bytes) attack.}
        \label{fig:class_lowH_rep-bytes}
\end{figure*}

In \Cref{img:low-H} and \Cref{img:rep-bytes}, we provide a more in-depth analysis of the performance obtained by our countermeasures against various forged files, named according to the notation described in \Cref{sec:dataset}. The plot in \Cref{img:low-H} shows the results of our mitigations against the files resulting from the low-H attack, while the plot in \Cref{img:rep-bytes} reports the results on the files forged with the rep-bytes attack. As previously discussed, since the com-seq attack is not consistently effective, we do not cover that attack. 

Both \Cref{img:low-H} and \Cref{img:rep-bytes} show that 2F is slightly better than the other countermeasures on most forged files, especially the files encrypted by NetWalker (NAR) and NotPetya (NEA) ransomware. For this reason, we have limited our investigation to 2F, 3F, and 4F. Increasing the number of fragments does not yield better results, and it has the downside of increasing linearly the time execution of the countermeasure. Among the various results, NetWalker enriched with l-h-e and rep-bytes is the ransomware that best avoids detection, keeping the accuracy of 3F and 4F to just over 60\%, and 70\% with 2F. On the other hand, other ransomware strains have a hard time escaping our countermeasures. For example, Sodinokibi (SOI) enriched with the rep-bytes attack cannot escape any of our countermeasures, with 3F and 4F hitting the 100\% accuracy mark, and 2F reaching 98\%.

\subsection{Mitigations Against Original Dataset}\label{sec:mit-original}
In the experiments described in \Cref{sec:mit-attack}, we have shown that our countermeasures are effective against the attacks that we proposed in \Cref{sec:attacks}. However, for our countermeasures to be usable in a realistic scenario, it is also important to verify that they are still able to identify files encrypted by in-the-wild ransomware. To do so, we have conducted another set of experiments, where we test our three countermeasures 2F, 3F, and 4F, on Davies et al. original dataset. To ensure that we test the robustness of our countermeasures on both datasets, in these experiments we have set the same parameters as the ones used in \Cref{sec:mit-attack}: distance, threshold, and subfragment length are the same in \Cref{table:res-org} and in \Cref{table:res-mix}.

As shown in \Cref{table:res-org}, when our mitigations are tested on the original dataset, the performance decreases, compared to the performance obtained by DAA. As visible in \Cref{table:res-org}, with respect to DAA accuracy result (98.24\%), 3F accuracy decreases by around 10\%, while for 2F and 4F it decreases by 7\%. Although the detection performance decreases, 2F, 3F, and 4F score an approximate accuracy of 90\%, confirming that they are able to retain efficacy against in-the-wild ransomware.

\begin{table}[t]
\centering
\normalsize
\caption{Best results of 2F, 3F and 4F countermeasures on the original dataset.}
\label{table:res-org}
\renewcommand\arraystretch{1.5}
\begin{tabular}{llll}
Data & 2F & 3F & 4F \\ 
\toprule
Length & 48 & 48 & 48\\
Threshold & 12 & 10 & 14\\
Distance & 54 & 48 & 56\\
Accuracy (\%) & 90.72 & 89.86 & \textbf{91.51}\\
Precision (\%) & 80.78 & 78.26 & \textbf{89.97}\\
Recall (\%) & 92.00 & 93.22 & \textbf{97.92}\\
F1 (\%) & 86.03 & 85.09 & \textbf{93.56}\\
\bottomrule
\end{tabular}
\end{table}

Regarding the other metrics, \Cref{table:res-org} shows that also our countermeasures take a performance hit for what it concerns the precision with respect to DAA. 4F is the one that obtains the best result (about 10\% lower than DAA), followed by 2F that shows a precision of about 20\% lower than DAA. Analysing the recall, 4F outperforms DAA with a 97.92\% result, versus 95\% of DAA. 2F and 3F follow close, with recall values of 92\% and 93.22\% respectively. In general, every method shows a recall above 92\%, proving that all approaches are able to keep low numbers of false negatives at identification time. Last, but not least, F1 scores for 2F reaches 86.03\%, while on 4F achieves a 93.56\%; in comparison, DAA obtained a 97.10\% F1 result.

\begin{figure*}
  \centering \includegraphics[width=0.95\linewidth]{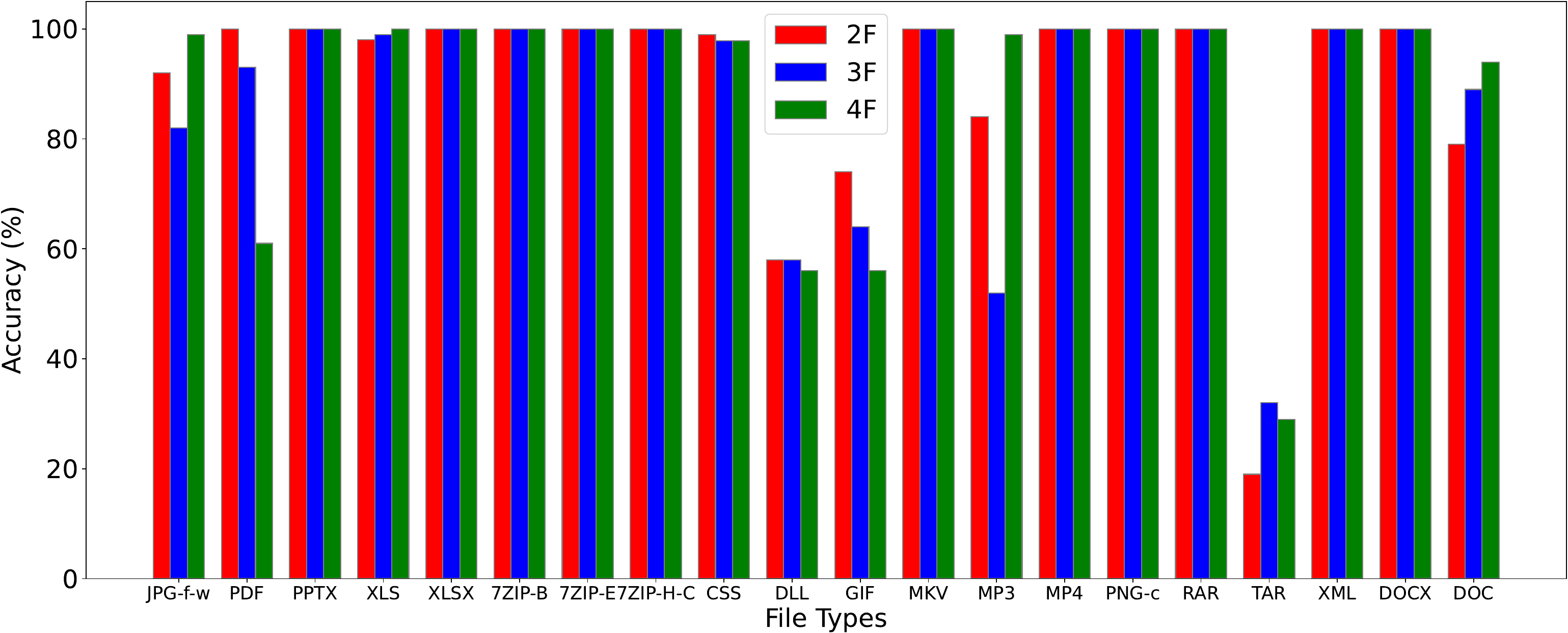}
  \caption{Overall classification accuracy of 2F, 3F and 4F for each common file type in the original dataset.}
  \label{img:acc-org-common}
\end{figure*}

\begin{figure*}
  \centering \includegraphics[width=0.95\linewidth]{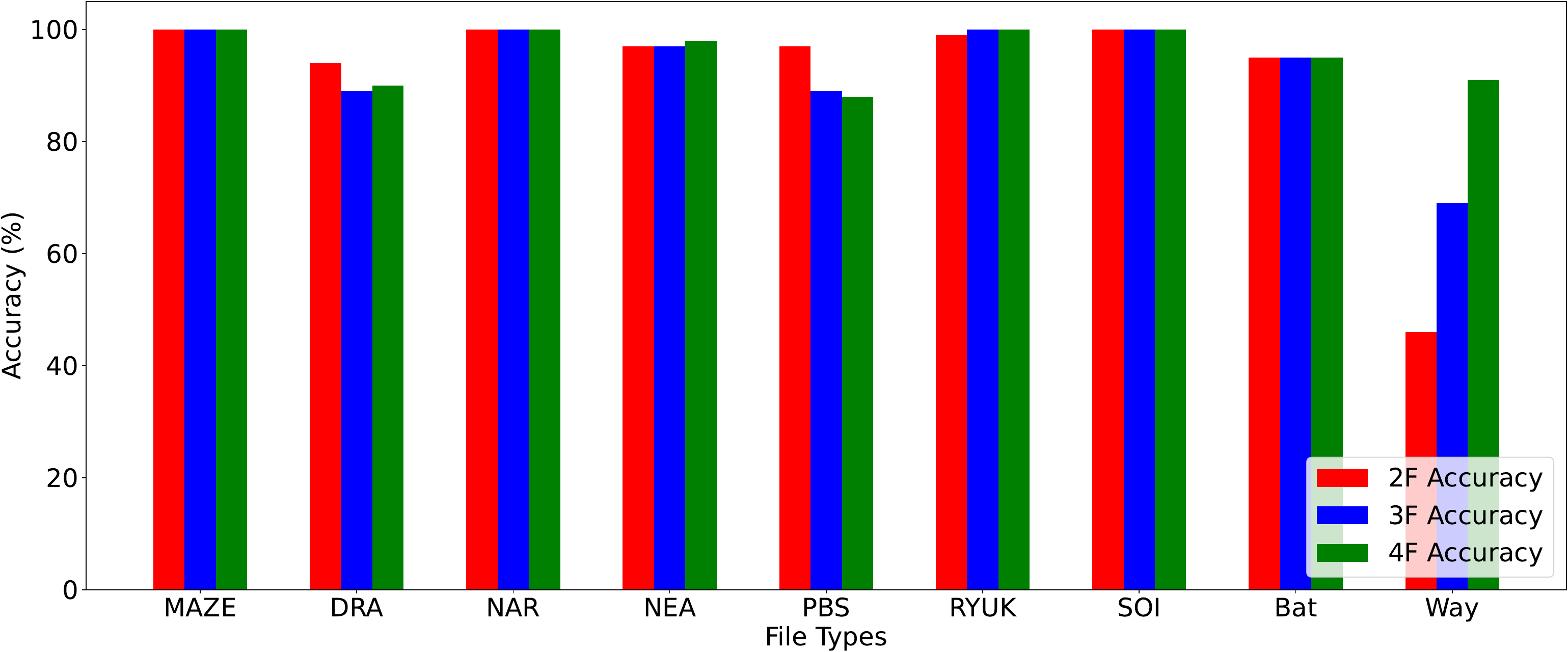}
  \caption{Overall classification accuracy of 2F, 3F and 4F for each ransomware file type in the original dataset.}
  \label{img:acc-org-ransom}
\end{figure*}

In \Cref{img:acc-org-common} and \Cref{img:acc-org-ransom}, we show a more in-detail overview of the performance of our countermeasures on the original dataset, with respect to the different file types we used. First, although the best countermeasure on the original dataset is 4F, the plots show that every countermeasure exhibits an accuracy higher than 88\% for the majority of the file types. By looking at the graphs, we can notice that there are some file types that challenge every countermeasure: DLLs, GIFs, TARs, and the WannaCry-encrypted files.

Our hypothesis is that all of these outlier results can be attributed to the specific entropy profiles of these file types. For example, DLLs contain a binary code part which exhibit high entropy, and bytes encoded with the LZW lossless data compression technique that exhibits high entropy as well. On a similar note, GIF images are compressed with the same LZW algorithm, which explains the relatively high entropy they exhibit, with respect to other file types.

Besides, it is possible to notice some differences in performance across countermeasures, depending on the file types. In particular, 4F performs noticeably better than 2F and 3F for identifying files encrypted by WannaCry, as well as for identifying legitimate JPGs, MP3s, and DOCs. However, 2F performs better when PDFs, GIF, Dharma, and Phobos are involved, while on TAR files it obtains the worst accuracy. The reason behind these results could be explained by the specific entropy-profile of these file types, with respect to other file extensions in the same category (i.e., "ransomware" or "non-ransomware" categories). In the case of WannaCry-encrypted files and DOC files, the more random fragments are used for the analysis, the better the accuracy. Our hypothesis, is that WannaCry produces encrypted files with a header entropy that is relatively low, with respect to other ransomware. This makes these files closer to legitimate files, entropy-wise, and harder to recognise if only the header is analysed. Similarly, DOC files might produce files with a relatively higher entropy in their header, with respect to other legitimate files, making it necessary to analyse more fragments to discern them from potential ransomware-encrypted files. 

The opposite reasoning applies to PDFs, GIFs, and Phobos-encrypted files, in which the fewer fragments we consider, the better the accuracy obtained. Our hypothesis is that the header entropy profile aligns with the other file types in their category, but that the entropy of the information contained in the body goes against the average entropy within-category (i.e., is noticeably lower or higher). This entails that the inclusion of additional fragments not only is not helpful, but even detrimental for the detection performance.

JPG, Dharma-encrypted, TAR, and MP3 files are peculiar, as they do not show a monotonic evolution with respect to the number of fragments used for the analysis. Taking JPEG files as an example, we can see that 2F performs quite well, while 3F performs worse, but then 4F performs best. For these file extensions, we presume that the random selection of the fragments have a considerable impact on the results obtained. Further studies would be necessary, to tailor our work specifically on these file types, adapting our countermeasures to their peculiar characteristics. 

The results on TAR files are also interesting. TAR files concatenate various files, and precede each of them with a header (containing file name, file mode, file size, etc.). Consequently, the entropy of the header fragment of a TAR file is low, if computed on the first 40 bytes, as Davies et al.~\cite{davies} describe. This implies that DAA correctly classifies a lot of TAR files, when using a proper threshold. Our countermeasures are less successful, due to two unfavourable factors. First, our countermeasures select multiple fragments and compute the entropy on each fragment, independently. Second, although TAR does not directly support data compression, it produces bigger archives than the ones obtained with other algorithms (e.g., LZW). These two factors make so that our countermeasures face a higher probability of selecting bad fragments, compute a higher total entropy than the one computed by DAA, and misclassify TAR files.

\subsection{Time Performance}\label{sec:time_perf}
Regarding our detection algorithms, it is important to take into account their performance in terms of required running time. For ransomware detection to be useful, not only it should be accurate, but also fast enough for minimising the number of maliciously infected files. Therefore, we have analysed the speed of our countermeasures against the speed of DAA to ensure that our modifications would not impact too negatively the execution time. 

In \Cref{table:time}, we report the time performance (in terms of files analysed per second) of DAA, 2F, 3F, and 4F, against our attack dataset. For what it concerns the parameters, for each countermeasure we use the best parameters that we identified in \Cref{sec:mit-attack}. As expected, the analysis speed decreases with the number of fragments; our data shows that DAA is the fastest of all algorithms with a performance of 50.22 files per second, while the slowest algorithm is 4F with 46.47 files per second. Although 2F, 3F, and 4F perform slightly worse than DAA, their speed is still sufficient for making them suitable detection algorithms. Taking into account the decreasing performance of 4F over 2F when dealing with the attack dataset, as presented in \Cref{table:daa-attacked}, and the fact that more fragments entail slightly fewer files analysed per second, introducing more fragments seems to have limited benefits.

Let us reason on the amount of data we need to analyse with our countermeasures, taking 4F as an example. The 4F algorithm, as shown in \Cref{table:res-org}, best performs with fragments length of $48$ bytes, meaning that running the 4F algorithm over one file requires the analysis of only $672$ bytes. In light of the resources available in a modern machine, analysing $672$ bytes produces a negligible overhead on the system. In a real scenario, to further minimise the impact on the system, it would be possible to trigger our algorithms on only recently modified files, after noticing a heavy hard-disk I\textbackslash O workload. Intuitively, these two aspect make the deployment of our countermeasures feasible.

\begin{table}[t]
\centering
\normalsize
\renewcommand\arraystretch{1.5}
\caption{Execution times (files per second) of DAA and our countermeasures on the attack dataset.}
\label{table:time}
\begin{tabular}{ll}
Algorithm & Files per Second\\ \toprule
DAA & 50.22\\
2F & 48.83\\
3F & 47.48\\
4F & 46.47 \\
\bottomrule
\end{tabular}
\end{table}

\subsection{Stress Test}\label{par:stress}
By their nature, entropy-based systems for ransomware detection can always be fooled by adding enough low-entropy data to a file.
For this reason, it is important to evaluate the amount of file manipulation that entropy-based detection approaches can sustain before becoming ineffective. In this section, we discuss a set of stress tests that we performed on our countermeasures 2F, 3F, and 4F, to verify how many bytes an attacker would have to add to circumvent detection.

First, as done in \Cref{sec:time_perf} for the time performance analysis, to test 2F, 3F, and 4F robustness, we used the parameters previously identified in \Cref{sec:mit-attack}. Then, for each of the 9 ransomware strains in our attack dataset, we modified the first 15 files in such a way that their entropy is lowered (as described in the following paragraph). Finally, we evaluated the accuracy that our countermeasures can still guarantee against these 135 forged files.

\begin{figure}[t]
\centering
\includegraphics[width=\linewidth]{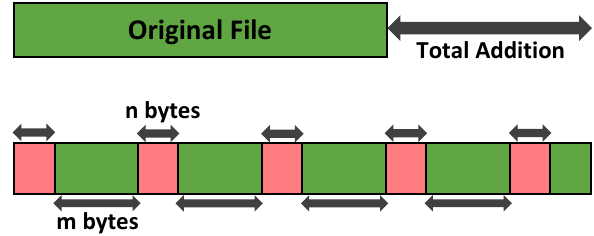}
\caption{Graphical representation of the stress tests conducted. In red, \textit{n} low entropy bytes are added every green \textit{m} bytes, until the end of the file.}
\label{img:stress-test}
\end{figure}

\Cref{img:stress-test} shows an overview on how we modified the files: every \textit{m} bytes of a file (that we named \textit{jump length}) we add \textit{n} bytes that contain a low entropy sequence.
The low-entropy sequences of \textit{n} bytes are created as follows:
\begin{itemize}
    \item A random byte is generated and repeated by steps of 2, from 0 up to 64 times, depending on the desired \textit{n} length;
    \item The sequence is inserted in each file every 4 to 64 bytes, by 4 bytes steps, until the end-of-file is reached.
\end{itemize}

It is worth noting that the modification described has a considerable impact on files composition, making it easy to detect when other features apart from entropy are used (e.g., I/O file buffer, or file size). While this approach is not suitable for avoiding detection, it is appropriate for evaluating the robustness of our countermeasures, for what it concerns the sole analysis of entropy.

Our experiments show that the countermeasure most affected by such modification is 3F, followed by 2F and 4F, in order. As expected, for every countermeasure, the more low-entropy sequences are injected in a file, the lower the detection performance is. This result is coherent with the results obtained during the analysis of our countermeasures applied to the original dataset \Cref{table:res-org}. There are two reasons why 3F performs worse than the other two mitigations. First, the optimal threshold and distance that we found for 3F are lower than the ones for 2F and 4F. Second, 3F scored the worst precision of all three mitigations. This is a direct consequence of aiming for the best accuracy possible, which requires balancing the parameters taking into consideration also the uninfected files.

\begin{figure}[tpb]
\captionsetup[subfigure]{aboveskip=2pt,belowskip=4pt}
     \begin{subfigure}[t]{\columnwidth}
         \includegraphics[width=\columnwidth]{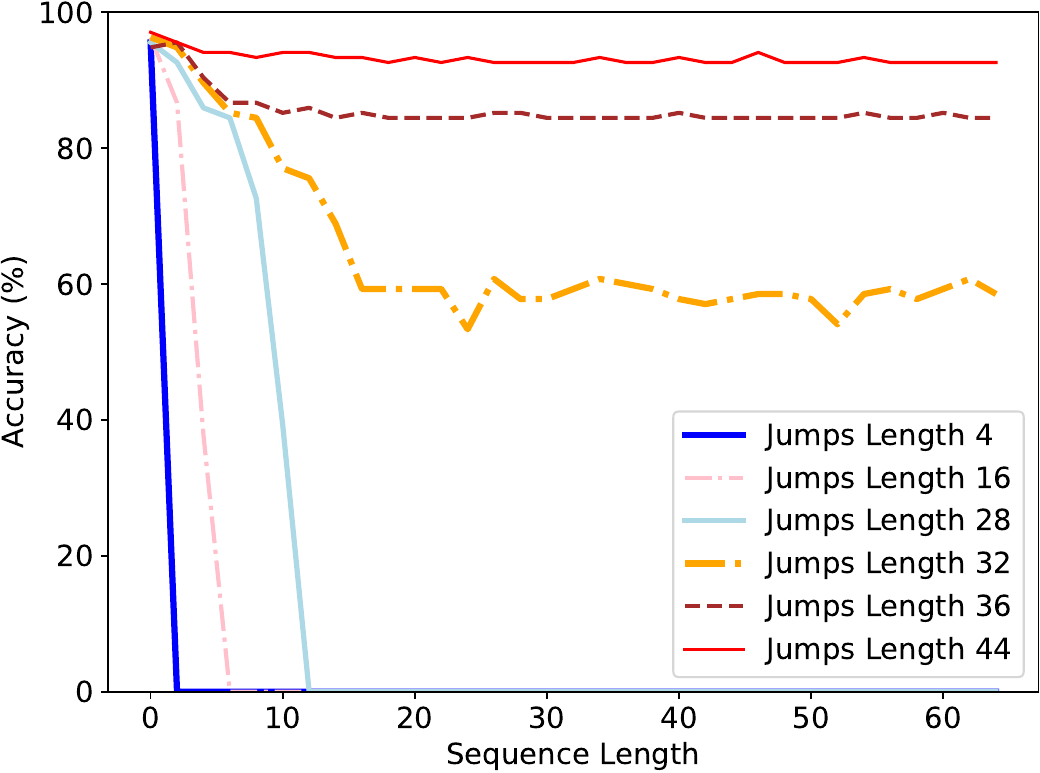}
         \caption{2F countermeasure}
         \label{img:s-test-2f}
     \end{subfigure}
     \begin{subfigure}[t]{\columnwidth}
         \includegraphics[width=\columnwidth]{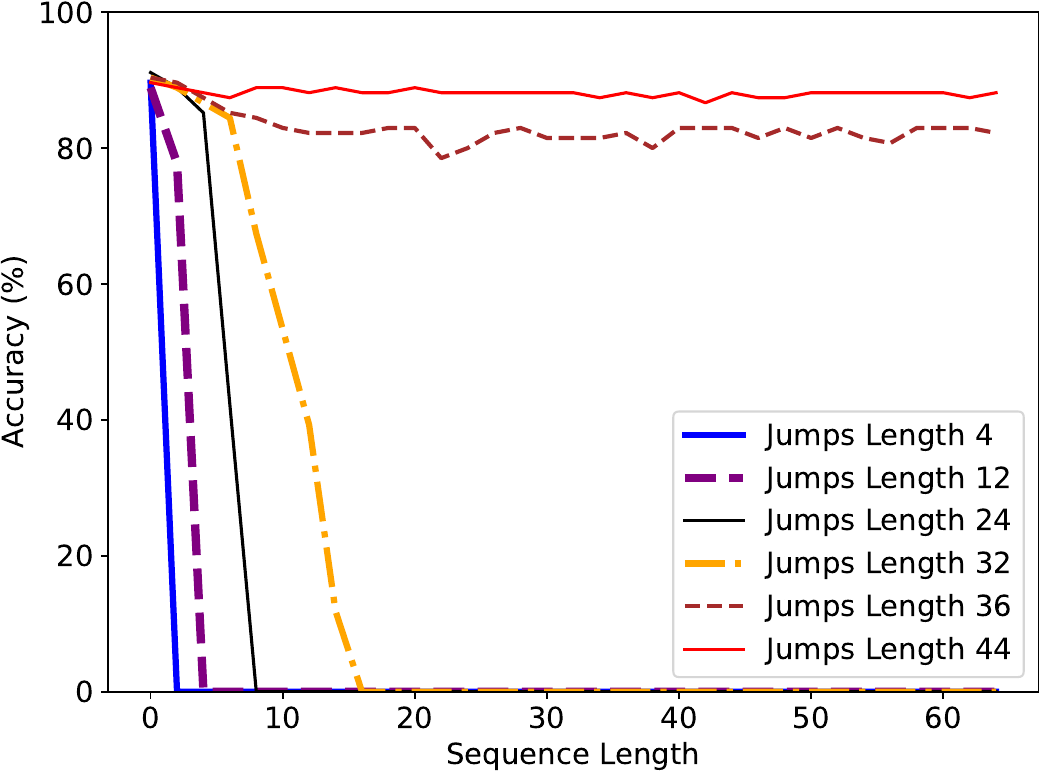}
         \caption{3F countermeasure}
         \label{img:s-test-3f-avg}
     \end{subfigure}
     \begin{subfigure}[t]{\columnwidth}
         \includegraphics[width=\columnwidth]{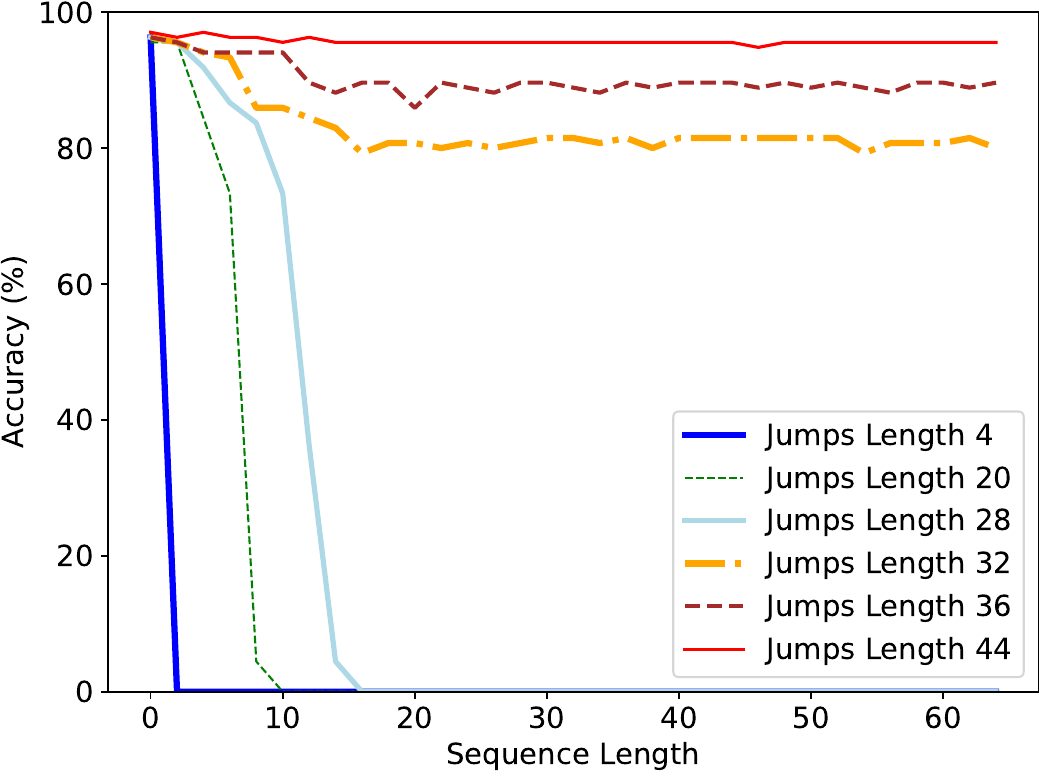}
         \caption{4F countermeasure}
         \label{img:s-test-4f-avg}
     \end{subfigure}
        \caption{Detection accuracy for each countermeasure, after stress test manipulations.}
        \label{fig:stress_tests_results}
\end{figure}

As shown in \Cref{img:s-test-3f-avg}, if an attacker aims to reduce 3F detection accuracy to 90\%, they should insert 2 bytes for every 44 bytes of the file, totalling for a 4.5\% of file size increase. \Cref{img:s-test-2f} shows that, for achieving the same result against 2F, an attacker should inject at least 8 bytes every 40 bytes, for a total of a 20\% size increase. Last, 4F proves to be the most resilient, requiring an attacker to inject 12 bytes of low-entropy data for every 36 bytes (\Cref{img:s-test-4f-avg}), which would lead to a 33\% file increase. In conclusion, 3F is the most susceptible to sequence injection, followed by 2F and 4F. 

\section{Limitations and Future Work}\label{sec:limitations}
In this paper, due to time and resource constraints, we had to heavily reduce the dataset size with respect to the one provided by Davies et al.~\cite{davies}. While their dataset contains around 1 million files, our dataset is composed of 2900 files. Although the size difference is noticeable, 
Davies et al.'s DAA performance on our reduced dataset is almost identical to the performance obtained on their larger dataset, showing that the composition of our dataset allows for an acceptable methodology comparison. In our future work, we will expand the dataset to include more files. On a similar note, our experiments showed that TAR, DLL, and GIF files exhibit strong resistance to our countermeasures. Again, increasing the dataset size might provide us more information to design ad-hoc solutions for such files.

It is worth noting that, in principle, our entropy-based approach is vulnerable to entropy manipulation, as any other entropy-based detection. However, due to the analysis of multiple fragments chosen at random positions within a file, ransomware would have to perform a considerable amount of modifications to escape our countermeasures. This would result in a significant file size increase, which can be detected by other methods running alongside ours.

In general, using encryption as an indicator of an ongoing ransomware attack could be problematic, as it could raise false alarms in a real-world scenario. For example, regular programs might perform encryption operations, and such operations could be flagged as malicious. However, due to its nature, ransomware encrypts a large amount of data as fast as possible, differently from benign software, which typically only operates on a limited number of files. Therefore, as we described in \Cref{sec:threat-model} and \Cref{sec:time_perf}, one possible solution would be to take into account the hard-disk I\textbackslash O rates, and program the real-time monitor to invoke our countermeasures only in case of high rates of writing events. Another workaround would be to equip the monitor with a pause function that allows users to suspend the monitor, before encrypting their files. This would avoid entirely the risk of triggering false positive warnings. 

DAA shares our same open questions, concerning the practical deployment of the detection algorithms themselves. Future work should focus on investigating these important aspects, related to the real-time monitor implementation, but out of scope of this paper, which focuses only on the ransomware detection algorithms.

Another small disadvantage of our algorithms and the one proposed by Davies et al.~\cite{davies} is that it activates during the mass-encryption phase, that is, after delivery and activation. At the beginning of this phase, ransomware already starts to alter files, and with our countermeasures the first attacked files cannot be spared from malicious encryption. For this reason, as  discussed in \Cref{sec:threat-model}, we recommend using our security countermeasures jointly with other defences, designed to identify ransomware at earlier infection stages. In this way, our countermeasures act as a last resort defence, attempting to limit the number of files impacted by ransomware that escaped upstream defences.

Finally, in this paper, our goal is to test the principal inner workings of our proposed attacks and countermeasures. In \Cref{sec:time_perf}, we have provided some discussion regarding the limited resources required for analysing a single file, independently of its size, and evaluation of the speed of our countermeasures. However, we do not conduct an in-depth analysis of potential side effects and overhead on system resources. Solid experiments in this regard, would require various tests on different hardware and OS configurations, making it intractable in the scope of this single paper. Further studies are warranted.

\section{Conclusion}\label{sec:conclusion}
In this paper, we have summarised Davies et al.~\cite{davies} DAA technique for detecting ransomware-encrypted files, and we have proposed three different attacks to lower its performance. Then, we have proposed three countermeasures (2F, 3F, and 4F) to mitigate our own attacks, and we have analysed their performance in terms of detection accuracy, time expenditure, and robustness. 
Our experiments prove that our attacks are effective at reducing the detection accuracy of DAA, and that our countermeasures are capable of identifying ransomware-encrypted files undergoing the effects of our attacks. We have also shown that the countermeasures are effective at identifying currently widespread ransomware, even though they incur in a performance penalty with respect to DAA, that goes from 7\% to 10\%. Of our countermeasures, 4F is the best performing across the board, taking into consideration every metric: accuracy, precision, F1, and recall.
\balance

Another interesting result is that all our countermeasures fail to classify correctly many TAR files, with a 35\% accuracy in the best case. Other file types, such as DLLs and GIFs, proved to be problematic for our countermeasures, with accuracy results ranging around 50\% - 60\% for DLLs and 55\% - 65\% for GIFs. In terms of time performance, our countermeasures fall short behind DAA, with time penalties that go from a mere 3\% in the case of 2F to 8\% for 4F. Last, but not least, we have shown that our countermeasures offer different levels of resilience against injection attacks of low-entropy sequences. To lower the detection accuracy of 3F below 90\%, an attacker would have to inject data that amounts to a 4.5\% file size increase. For achieving the same result with 4F, an attacker would have to inject enough data for increasing the file size by 33\%. In conclusion, in this paper we have highlighted that entropy-based detection systems can be circumvented with careful manipulation of files entropy. However, we have also proven that entropy still has merits as a detection feature, if properly used.

\bibliographystyle{IEEEtran}
\bibliography{IEEEabrv,./bibliography.bib}

\begin{thebibliography}{10}
\providecommand{\url}[1]{#1}
\csname url@samestyle\endcsname
\providecommand{\newblock}{\relax}
\providecommand{\bibinfo}[2]{#2}
\providecommand{\BIBentrySTDinterwordspacing}{\spaceskip=0pt\relax}
\providecommand{\BIBentryALTinterwordstretchfactor}{4}
\providecommand{\BIBentryALTinterwordspacing}{\spaceskip=\fontdimen2\font plus
\BIBentryALTinterwordstretchfactor\fontdimen3\font minus \fontdimen4\font\relax}
\providecommand{\BIBforeignlanguage}[2]{{%
\expandafter\ifx\csname l@#1\endcsname\relax
\typeout{** WARNING: IEEEtran.bst: No hyphenation pattern has been}%
\typeout{** loaded for the language `#1'. Using the pattern for}%
\typeout{** the default language instead.}%
\else
\language=\csname l@#1\endcsname
\fi
#2}}
\providecommand{\BIBdecl}{\relax}
\BIBdecl

\bibitem{sonicwall2022}
\BIBentryALTinterwordspacing
``2022 sonicwall cyber threat report,'' 2022. [Online]. Available: \url{https://www.sonicwall.com/2022-cyber-threat-report/}
\BIBentrySTDinterwordspacing

\bibitem{statistafamilies}
\BIBentryALTinterwordspacing
``Number of newly discovered ransomware families worldwide from 2015 to 2020,'' 2021. [Online]. Available: \url{https://www.statista.com/statistics/701029/number-of-newly-added-ransomware-families-worldwide/}
\BIBentrySTDinterwordspacing

\bibitem{statistaattacks}
\BIBentryALTinterwordspacing
``Number of newly discovered ransomware families worldwide from 2015 to 2020,'' 2021. [Online]. Available: \url{https://www.statista.com/statistics/494947/ransomware-attacks-per-year-worldwide/}
\BIBentrySTDinterwordspacing

\bibitem{portswigger}
\BIBentryALTinterwordspacing
``Dax-côte d’argent hospital in france hit by ransomware attack,'' 2021. [Online]. Available: \url{https://portswigger.net/daily-swig/dax-cote-dargent-hospital-in-france-hit-by-ransomware-attack}
\BIBentrySTDinterwordspacing

\bibitem{alrimy2018}
B.~A.~S. Al-rimy, M.~A. Maarof, and S.~Z.~M. Shaid, ``Ransomware threat success factors, taxonomy, and countermeasures: A survey and research directions,'' \emph{Computers \& Security}, vol.~74, pp. 144--166, 2018.

\bibitem{kharraz2015}
A.~Kharraz, W.~Robertson, D.~Balzarotti, L.~Bilge, and E.~Kirda, ``Cutting the gordian knot: A look under the hood of ransomware attacks,'' in \emph{Detection of Intrusions and Malware, and Vulnerability Assessment}, M.~Almgren, V.~Gulisano, and F.~Maggi, Eds.\hskip 1em plus 0.5em minus 0.4em\relax Cham: Springer International Publishing, 2015, pp. 3--24.

\bibitem{davies}
S.~R. Davies, R.~Macfarlane, and W.~J. Buchanan, ``Differential area analysis for ransomware attack detection within mixed file datasets,'' \emph{Computers \& Security}, vol. 108, p. 102377, 2021.

\bibitem{shannon1948}
C.~E. Shannon, ``A mathematical theory of communication,'' \emph{The Bell System Technical Journal}, vol.~27, no.~3, pp. 379--423, 1948.

\bibitem{zhu2022}
J.~Zhu, J.~Jang-Jaccard, A.~Singh, I.~Welch, H.~AI-Sahaf, and S.~Camtepe, ``A few-shot meta-learning based siamese neural network using entropy features for ransomware classification,'' \emph{Computers \& Security}, vol. 117, p. 102691, 2022.

\bibitem{hirano2019}
M.~Hirano and R.~Kobayashi, ``Machine learning based ransomware detection using storage access patterns obtained from live-forensic hypervisor,'' in \emph{2019 Sixth International Conference on Internet of Things: Systems, Management and Security (IOTSMS)}, 2019, pp. 1--6.

\bibitem{cuzzocrea2018}
A.~Cuzzocrea, F.~Martinelli, and F.~Mercaldo, ``A novel structural-entropy-based classification technique for supporting android ransomware detection and analysis,'' in \emph{2018 IEEE International Conference on Fuzzy Systems (FUZZ-IEEE)}, 2018, pp. 1--7.

\bibitem{continella}
A.~Continella, A.~Guagnelli, G.~Zingaro, G.~De~Pasquale, A.~Barenghi, S.~Zanero, and F.~Maggi, ``Shieldfs: A self-healing, ransomware-aware filesystem,'' in \emph{Proceedings of the 32nd Annual Conference on Computer Security Applications}, ser. ACSAC '16.\hskip 1em plus 0.5em minus 0.4em\relax New York, NY, USA: Association for Computing Machinery, 2016, p. 336–347.

\bibitem{genc2018}
Z.~A. Gen{\c{c}}, G.~Lenzini, and P.~Y.~A. Ryan, ``Next generation cryptographic ransomware,'' in \emph{Secure IT Systems}, N.~Gruschka, Ed.\hskip 1em plus 0.5em minus 0.4em\relax Cham: Springer International Publishing, 2018, pp. 385--401.

\bibitem{zhao}
B.~Zhao, Q.~Liu, and X.~Liu, ``Evaluation of encrypted data identification methods based on randomness test,'' in \emph{2011 IEEE/ACM International Conference on Green Computing and Communications}, 2011, pp. 200--205.

\bibitem{alekseev}
I.~Alekseev and V.~Platonov, ``Detection of encrypted executable files based on entropy analysis to determine the randomness measure of byte sequences,'' \emph{Automatic Control and Computer Sciences}, vol.~51, no.~8, pp. 915--920, 2017.

\bibitem{scaife}
N.~Scaife, H.~Carter, P.~Traynor, and K.~R.~B. Butler, ``Cryptolock (and drop it): Stopping ransomware attacks on user data,'' in \emph{2016 IEEE 36th International Conference on Distributed Computing Systems (ICDCS)}, 2016, pp. 303--312.

\bibitem{jung}
S.~Jung and Y.~Won, ``Ransomware detection method based on context-aware entropy analysis,'' \emph{Soft Computing}, vol.~22, no.~20, pp. 6731--6740, 2018.

\bibitem{lee}
K.~Lee, S.-Y. Lee, and K.~Yim, ``Machine learning based file entropy analysis for ransomware detection in backup systems,'' \emph{IEEE Access}, vol.~7, pp. 110\,205--110\,215, 2019.

\bibitem{hsu2021}
C.-M. Hsu, C.-C. Yang, H.-H. Cheng, P.~E. Setiasabda, and J.-S. Leu, ``Enhancing file entropy analysis to improve machine learning detection rate of ransomware,'' \emph{IEEE Access}, vol.~9, pp. 138\,345--138\,351, 2021.

\bibitem{mcintosh19}
T.~McIntosh, J.~Jang-Jaccard, P.~Watters, and T.~Susnjak, ``The inadequacy of entropy-based ransomware detection,'' in \emph{Neural Information Processing}, T.~Gedeon, K.~W. Wong, and M.~Lee, Eds.\hskip 1em plus 0.5em minus 0.4em\relax Cham: Springer International Publishing, 2019, pp. 181--189.

\bibitem{pont}
J.~Pont, B.~Arief, and J.~Hernandez-Castro, ``Why current statistical approaches to ransomware detection fail,'' in \emph{Information Security: 23rd International Conference, ISC 2020, Bali, Indonesia, December 16–18, 2020, Proceedings}.\hskip 1em plus 0.5em minus 0.4em\relax Berlin, Heidelberg: Springer-Verlag, 2020, p. 199–216.

\bibitem{atkinson}
K.~E. Atkinson, \emph{An introduction to numerical analysis}.\hskip 1em plus 0.5em minus 0.4em\relax John wiley \& sons, 2008.

\bibitem{github}
\BIBentryALTinterwordspacing
Gomitolof, ``Gomitolof/daa-for-ransomware-detection: Ransomware detection using shannon entropy and differential area analysis of the files.'' [Online]. Available: \url{https://github.com/gomitolof/DAA-for-Ransomware-detection}
\BIBentrySTDinterwordspacing

\end{thebibliography}
\end{document}